%% file: root.tex
\documentclass[journal]{IEEEtran}
\input{preambleT.tex}
\usepackage[colorlinks = true,
linkcolor = blue,
urlcolor  = blue,
citecolor = blue,
anchorcolor = blue]{hyperref}

\begin{document}
\title{Energy Crowdsourcing and Peer-to-Peer Energy Trading in Blockchain-Enabled Smart Grids}

\author{Shen Wang,~\IEEEmembership{Student Member,~IEEE,} Ahmad~F.~Taha,~\IEEEmembership{Member,~IEEE,} Jianhui~Wang,~\IEEEmembership{Senior Member,~IEEE,} \\
     Karla Kvaternik,~\IEEEmembership{Member,~IEEE,}
          Adam Hahn,~\IEEEmembership{Member,~IEEE.}
\thanks{S. Wang and A. F. Taha are with the Department
of Electrical and Computer Engineering, The University of Texas at San Antonio,
TX. J. Wang is with the Department of Electrical Engineering, Southern Methodist University, Dallas, TX.      Karla Kvaternik is with Siemens Corporate Technology, Princeton, NJ. Adam Hahn is with the School of Electrical Engineering and Computer Science, Washington State University, Pullman, WA. E-mails: \texttt{mvy292@my.utsa.edu, ahmad.taha@utsa.edu, jianhui@smu.edu, karla.kvaternik@siemens.com, ahahn@eecs.wsu.edu}. An earlier version of this paper was presented at the 2018 IEEE Power \& Energy Society General Meeting in Portland, Oregon, August 5--9, 2018. A preprint of the conference paper can be found in~\cite{wang2018blockchain}, which shows the significant extensions and contributions of this paper in comparison with~\cite{wang2018blockchain}.} }
\maketitle

\begin{abstract}
The power grid is rapidly transforming, and while recent grid innovations increased the utilization of advanced control methods, the next-generation grid demands technologies that enable the integration of distributed energy resources (DERs)---and consumers that both seamlessly buy and sell electricity. This paper develops an optimization model and blockchain-based architecture to manage the operation of crowdsourced energy systems (CES), with peer-to-peer (P2P) energy trading transactions. An operational model of CESs in distribution networks is presented considering various types of energy trading transactions and crowdsourcees. Then, a two-phase operation algorithm is presented: Phase I focuses on the day-ahead scheduling of generation and controllable DERs, whereas Phase II is developed for hour-ahead or real-time operation of distribution networks.  The developed approach supports seamless P2P energy trading between individual prosumers and/or the utility. The presented operational model can also be used to operate islanded microgrids.
The CES framework and the operation algorithm are then prototyped through an efficient blockchain implementation, namely the IBM Hyperledger Fabric. This implementation allows the system operator to manage the network users to seamlessly trade energy.  Case studies and prototype illustration are provided.
\end{abstract}

\begin{IEEEkeywords}
Energy Crowdsourcing, Blockchain, Energy Trading, Peer-to-Peer Energy Management.
\end{IEEEkeywords}
\section{Introduction}
\IEEEPARstart{S}{mart} grid technologies, such as microgrids and distributed energy resources (DERs), have drastically changed the way electricity is generated and consumed in two dimensions. First, the rapid increase in energy \textit{prosumers} introduces new grid participants and provides {a more decentralized and} {open} power grid. Second, this changes the role of a system operator or utility from a power retailer to a service provider---renting transmission/distribution lines to prosumers, rather than solely selling units of energy. This paradigm shift requires the creation of new trusted software platforms, distributed operation/control algorithms, and computational methods to enable reliable grid operations, prosumer engagement, and incentivize utility business model innovations. 

\begin{figure}[t]
	\centering
	\includegraphics[scale=0.32]{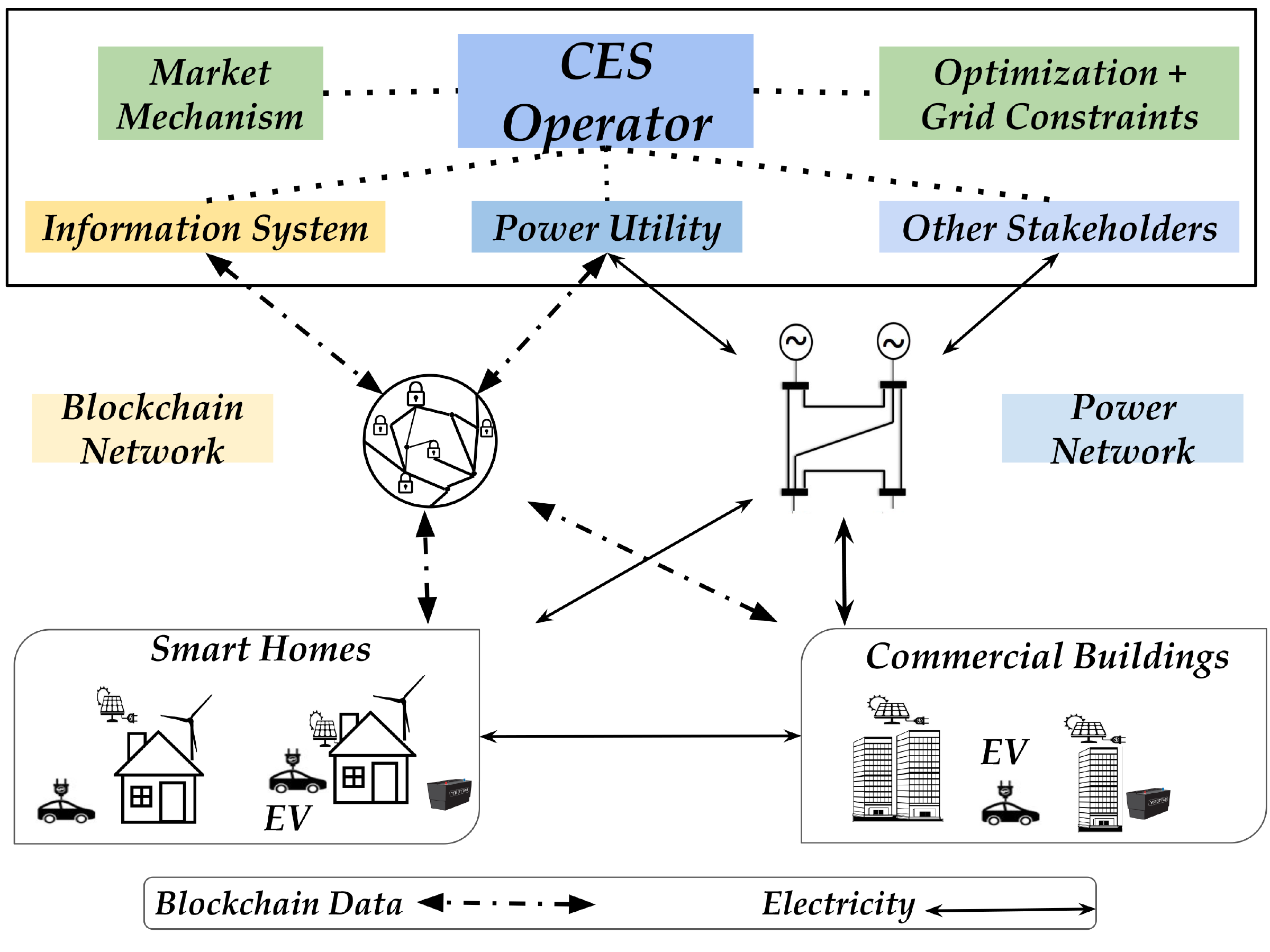}        
	\caption{Blockchain-assisted architecture of operation in CESs.}
	\label{fig:ces}
\end{figure}
Crowdsourcing~\cite{howe2006rise} is a major drive for various industries, and has been utilized in various disciplines such as medicine, cyber physical systems, and engineering system design. The central theme in crowdsourcing is the utilization of the crowd's power to achieve system-level objectives. To see how crowdsourcing can be applied in energy systems, we provide an analogy from the most popular crowdsourcing markets, the Amazon Mechanical Turk (MTurk)~\cite{ipeirotis2010analyzing}, which enables people to post jobs with monetary rewards and expiry dates.  Energy crowdsourcing offers the possibility of the transformation in energy systems, and this paper puts forth operational models of crowdsourced energy system for collaborative production and consumption in energy markets, shown in Fig.~\ref{fig:ces}. The tasks in crowdsourced energy system can be plugging in an electric vehicle, charging/discharging a battery, deferring loads, and supplying the power network with renewable energy via solar panels---with the objective of satisfying a near-real-time demand shortage/surplus. These tasks can be automated via smart inverters, plugs, and meters while interfacing with power utilities and a distributed blockchain implementation.

This transformation in sustainable energy systems, where energy management is crowdsourced by prosumers, will be supported by two key, disruptive scientific technologies: \textit{(i)} new modeling and crowdsourcing-centered methods that perform real-time grid management while maintaining the grid's stability. \textit{(ii)} A secure cyber-infrastructure design to manage and coordinate millions of energy-trading transactions (\textit{prosumer-prosumer} or \textit{prosumer-operator trades}).

The majority of the new modeling methods are based on optimal power flow (OPF) operation models and the secure cyber-infrastructure design is implemented by the promising blockchain technology. However, both the new modeling methods and the implementation of blockchain have limitations.  First, the computed OPF setpoints for DERs and controllable loads might not be eventually adopted by crowdsourcees and prosumers. Second, it is unclear how energy trading between prosumers can take place within the operational models. Third, the utilized blockchain architectures are not scalable to include millions of energy trading transactions---especially that blockchain-based trades consume a significant amount of energy. The paper addresses these gaps, and the main contributions and organization are given as follows.
\begin{itemize}
	\item An operational framework and model of crowdsourced energy systems in distribution networks is presented considering various types of energy trading transactions and crowdsourcees. The presented framework enables P2P energy trading at the distribution level, where ubiquitous distribution-level asset owners can trade with each other. This has not done before in association with distributed OPF routines and blockchain-enabled architecture.  In such a framework, an operator is needed to clear the market and ensure there is no violation of any technical constraints (e.g., distribution line limits). A distribution system operator can assume this role running the presented CES operational model (Section~\ref{sec:Model}). Extensions to operator-free, islanded microgrids are also showcased.
	\item A two-phase, near real-time operation algorithm for crowdsourced energy systems is explored. The first phase focusing on the day-ahead scheduling of generation and controllable DERs manages the bulk of grid-operation, while the second phase is developed to balance hour-ahead even real-time deficit/surplus in energy via monetary incentives. The developed two-phase algorithm supports arbitrary P2P energy trading between prosumers and utility, resulting in a systematic way to manage distribution networks amid P2P energy trading while incentivizing crowdsourcees to contribute to this ecosystem. The algorithm supports operation of islanded, self-autonomous microgrid (Section~\ref{sec:OPF-ID}).  
	\item The CES framework is implemented and prototyped within IBM Hyperledger Fabric platform---an efficient blockchain implementation. This implementation allows the system operator to manage the network and supports users to log in, manage their own account and carry on the energy trading with utilities or neighborhoods. This prototype communicates with the two-phase algorithm presented in this paper, is open source, and can be used by utilities (Section~\ref{sec:BlockChain}).  Finally, numerical tests on a distribution network and blockchain prototype illustration are provided (Section~\ref{sec:Case-Studies}).
\end{itemize}

\section{Literature Review}   ~\label{sec:Literature}
\vspace{-1cm}
\subsection{Grid Operation, OPF, and Demand Response}  ~\label{sec:operation}
Recent studies have investigated integrating the operation of DERs in distribution networks. The focus of majority of these studies~\cite{he2018second,He2018b} is on unit commitment, economic dispatch problems, scheduling of DERs, and maintaining the grid's frequency and voltage within acceptable ranges while given uncertainty from renewables and load forecasts.

Another branch of related work~\cite{hajiesmaili2017crowd} studies the design of demand response signals and incentives to drive DER owners to contribute to energy production. In summary, there are three approaches to demand response: \textit{(a)} Reducing demand by using local DERs. \textit{(b)} Reducing demand through shifting controllable loads. \textit{(c)} Designing efficient generator setpoints to reduce the total generation~\cite{DR1}. The majority of demand response schedules focus on operational timescale. Further, the need for real-time regulation and distributed dynamic pricing as a function of the grid's physical status motivates new physics-aware pricing mechanisms~\cite{langbort2010real,Namerikawa2015}.  Background on blockchain and energy trading routines is given next.
\begin{table}[t]
	\renewcommand{\arraystretch}{1.4}
	\fontsize{7.5}{6}\selectfont
	\caption{Various Implementations of blockchain. PoW and RBFT stand for Proof of Work and Redundant Byzantine Fault Tolerance.}
	\centering
	\begin{tabular}{ c|c|c|c }
		& \textit{Bitcoin} & \textit{Ethereum} & \textit{Hyperledger Fabric} \\
		\hline
		\textit{Cryptocurrency} & Bitcoin & Ether & None \\
		\hline
		\textit{Network}  & public & public & permissioned \\ 
		\hline
		\textit{Transactions} & anonymous & anonymous & public/confidential \\        
		\hline
		\textit{Consensus} & PoW & PoW & RBFT \\         
		\hline
		\textit{Smart Contracts}    & None & Solidity & Chaincode \\         
		\hline
		\textit{Language} &    C++ & C++/Golang &    Golang/Java \\
		\hline        \hline
	\end{tabular}
	\label{table:BlockchainType}
\end{table}
\subsection{Blockchain and Energy Trading Systems}  \label{sec:ETS} 
Blockchain is a distributed ledger based on a set of communication and consensus protocols that ensure the ledger integrity through interlinked, cryptographically signed, and time-stamped blocks that define transactions~\cite{Yuan2018}. 
	\begin{table*}[t]
		\centering	{	\renewcommand{\arraystretch}{1.5}
		\centering
		\begin{threeparttable}		\centering
			\caption{Various focuses of typical P2P energy trading systems.}
			\begin{tabular}{ c|c|c|c|c|c|c|c|c }
				\textit{Reference}  &  {\cite{munsing2017blockchains}}  & \cite{luo2018distributed} & \cite{su2018secure}  & \cite{kang2017enabling}  & \cite{paudel2018peer} & \cite{long2018peer} & \cite{huang2019optimal} & \cite{Hahn2017}\\
				\hline \textit{Market Mechanism}  &\pie{90} &\pie{90}& \pie{180}& \pie{150} & \pie{360} & \pie{270} & \pie{90} & \pie{270}\\
				\hline 
				\textit{Information System} &  Blockchain &  Blockchain& Blockchain  & Blockchain & \pie{0} &  \pie{0} & Blockchain &Blockchain\\
				\textit{(Consensus)} & PoW(Public) & Self-designed& Self-designed& NA & NA & NA & NA  & PoW\\
				\hline 
				\textit{\makecell{Optimization + \\ Grid Constraints}} & \pie{360}& \pie{360} & \pie{360}  & \pie{360} & \pie{90}& \pie{360} & \pie{360}& \pie{0}\\
				\hline 
				\textit{Scenario}&  Microgrid& EV & EV  & EV & Microgrid & Microgrid & EV & Microgrid\\
				\hline        \hline
			\end{tabular}
			\label{table:literature_review}
			\begin{tablenotes}
				\footnotesize
				\item \pie{0} means  \textit{not considered}; \pie{100} means \textit{partially considered}; \pie{360} means \textit{fully considered}; NA means \textit{not applicable} or the authors do not cover the aspect; \textit{Self-designed} means that authors design a new, corresponding  consensus mechanism for their own blockchain implementation; 
			\end{tablenotes}
		\end{threeparttable}
	}
\end{table*}
The blockchain concept originated with the Bitcoin protocol, which utilized a proof of work (PoW) consensus mechanism where miners combine transactions into Merkle tree-based blocks and compete to find a random nonce that
produces a hash digest within a predefined range. However, this approach has many limitations 
including its significant energy consumption, scalability in the number of transactions/seconds, privacy concerns with a public ledger, and single purpose application (i.e., an exchange of the Bitcoin cryptocurrency~\cite{swan2015blockchain}). A number of additional blockchain technologies have been introduced to address these challenges as suggested below:
\begin{itemize}
\item \textit{Efficient consensus mechanisms:} A consensus protocol is used to ensure the unambiguous ordering of transactions and guarantees the integrity and consistency of the blockchain across distributed nodes~\cite{baliga2017understanding};  
the annual estimated electricity consumption of Bitcoin PoW consensus is 47.1 Terawatt-hour---a staggering 0.21\% of worlds electricity consumption~\cite{energyconsumption}. Furthermore, PoW techniques typically have limitations on the number of transactions per second, which limits use in high performance environments. Other consensus mechanisms, such as Proof of Stake (e.g., Ethereum Casper~\cite{zamfir2015introducing})
 or Redundant Byzantine Fault Tolerance (RBFT) (e.g., IBM Hyperledger Fabric~\cite{hyperledger}), can be used to reduce energy consumption.
\item \textit{Smart contracts:} Smart contracts provide protocols and Turing complete virtual machines that enable nodes to execute some program based on the results of new transactions and allow the blockchain to support sophisticated logic. Smart contracts and blockchain provide an excellent platform to perform energy trading transactions. In particular, the authors in~\cite{licata_2017} provide a high-level description to the main merits of using cryptocurrency and blockchain in energy systems. 
\item \textit{Permissioned and privacy mechanisms:} Blockchain platforms can be categorized into public and private, where public implies that any miner can contribute to the consensus and block creation, while permissioned chains restrict block creation to a predefined set of parties. Therefore, permissioned chains may be preferred in applications with defined authorities or entities with management responsibilities.
\end{itemize}
Tab.~\ref{table:BlockchainType} summarizes the attributes of different implementations of current blockchains, and Section~\ref{sec:BlockChain} provides additional discussion on why the Hyperledger platform is selected to implement the proposed crowdsourced energy system scheme.

In Tab.~\ref{table:literature_review}, various focuses of recent P2P energy trading routines are compared according to focus aspects; the first three aspects are derived from~\cite{mengelkamp2018designing}. These aspects reflect corresponding modules in Fig.~\ref{fig:ces}, and are explained here. First, The \textit{market mechanism} including the participant setup, and pricing mechanism is designed to incentivize participants while maximizing the social welfare. The participant setup defines market participants, and the form of energy trading, while pricing mechanism, i.e., incentive design, and bidding strategy, comprises the market’s allocation and payment rules.  
Second, the \textit{information system}  is designed to connect all market participants, provide the market platform, offer market access, and monitor the market operations. Nowadays blockchain is suitable to implement part of \textit{information system}. Third, the \textit{optimization and grid constraints} refer to scheduling of DERs while maintaining the grid in an optimal way as we discussed in Section~\ref{sec:operation}. Our corresponding implementation of the above aspects are presented in Section~\ref{sec:Implementation}. Finally, as for the \textit{scenario} in these papers, we notice that most papers focuses on microgrids and electric vehicles.

After comparing the typical papers in Tab.~\ref{table:literature_review}, we notice the following. First, references \cite{munsing2017blockchains,luo2018distributed,su2018secure,kang2017enabling,huang2019optimal} focus more on approach to managing the grid with the assistance of simple negotiation, auction, or bidding mechanism and implementing the information system via thriving blockchain technology, since the security and privacy can be guaranteed. Specifically, The contribution of \cite{luo2018distributed} is more about the multi-agent system based trading negotiation mechanism. The authors in~\cite{su2018secure} propose a contract based blockchain for secure EV charging, and a reputation based Byzantine fault tolerance consensus algorithm is proposed. In \cite{huang2019optimal}, the new and hybrid charging scenario, i.e., mobile charging vehicle-to-vehicle, and grid-to-vehicle are considered.  Second, the authors in~\cite{paudel2018peer,long2018peer,Hahn2017} pay attention on designing different marketing/pricing mechanism, but the power flow model is ignored in their optimization. For example, game theoretical approaches are adopted to achieve real-time pricing in~\cite{paudel2018peer,wang2016reinforcement,park2016contribution}.
Besides the typical paper listed above, the attack/threat model are explored further in energy blockchain in \cite{gai2019privacy,zhu2019controllable} to enhance the security and privacy. Especially in \cite{zhu2019controllable}, the authors design a special trust authority node  with a veto power to prevent malicious voting. However, the marketing/pricing mechanism and platform design for P2P energy markets do not receive too much attention and still are an open research area.

Beyond research-oriented studies, companies (i.e., \cite{daisee,gridplus,bankymoon}) mainly focus on the development of business models, and the possibility of introducing those models to local energy market and design of control systems are not fully considered.
\section{Integrated Operational Model of CESs}~\label{sec:Model}
In this section, we present an integrated operational model of crowdsourced energy systems that considers a wide range of DERs, different types of crowdsourcees and energy trading transactions in distribution networks.
 For simplicity, we focus on radial distribution networks with a single feeder connected to traditional generation and utility-scale renewables.  
We consider a CES at the feeder level with $n$ buses modeled by a tree graph $(\mathcal{N},\mathcal{E})$, where $\mathcal{N}= \{1,\ldots, n\}$ is the set of nodes and $\mathcal{E} \subseteq \mathcal{N}\times\mathcal{N}$ is the set of lines. Define the partition $\mathcal{N} = \mathcal{G} \bigcup \mathcal{C} \bigcup \mathcal{L} $, where $\mathcal{G} = \{1,\ldots,n_g\}$ collects the $n_g$ utility-scale power generation connected to the feeder/substation; $\mathcal{C} = \{1,\ldots, n_c\}$ collects the buses containing $n_c$ users who signed up for crowdsourcing schedules; $\mathcal{L} = \{1,\ldots, n_l\}$ collects load buses.
\begin{figure}[t]
    \centering \includegraphics[scale=0.21]{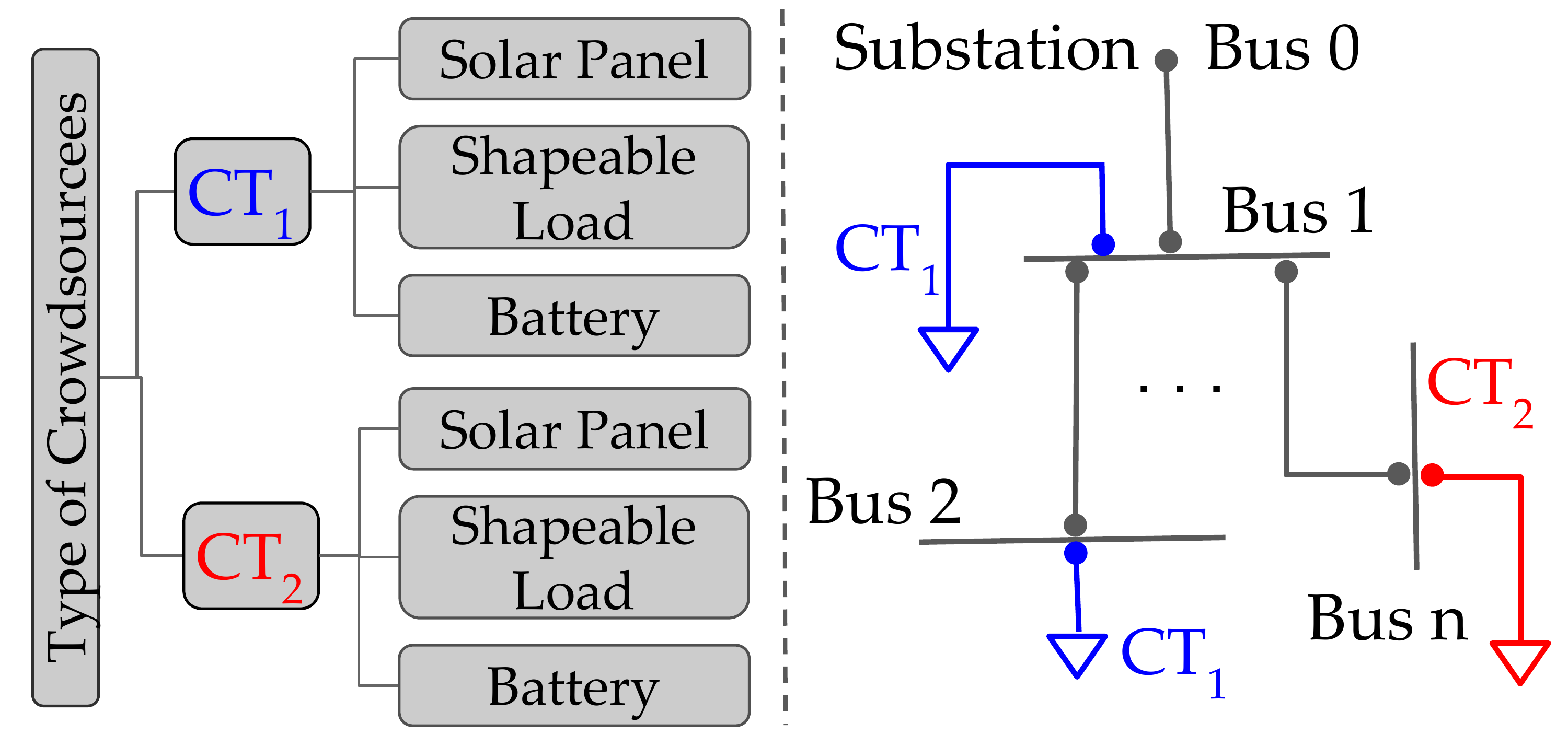}
    \caption{A radial network with different types of crowdsourcees: $\mc{CT}_{1}$ (blue) and $\mc{CT}_{2}$ (red).}
    \label{fig:UserType}
   \vspace{-0.5cm}
\end{figure}
\begin{figure}[t]
	\centering \includegraphics[scale=0.21]{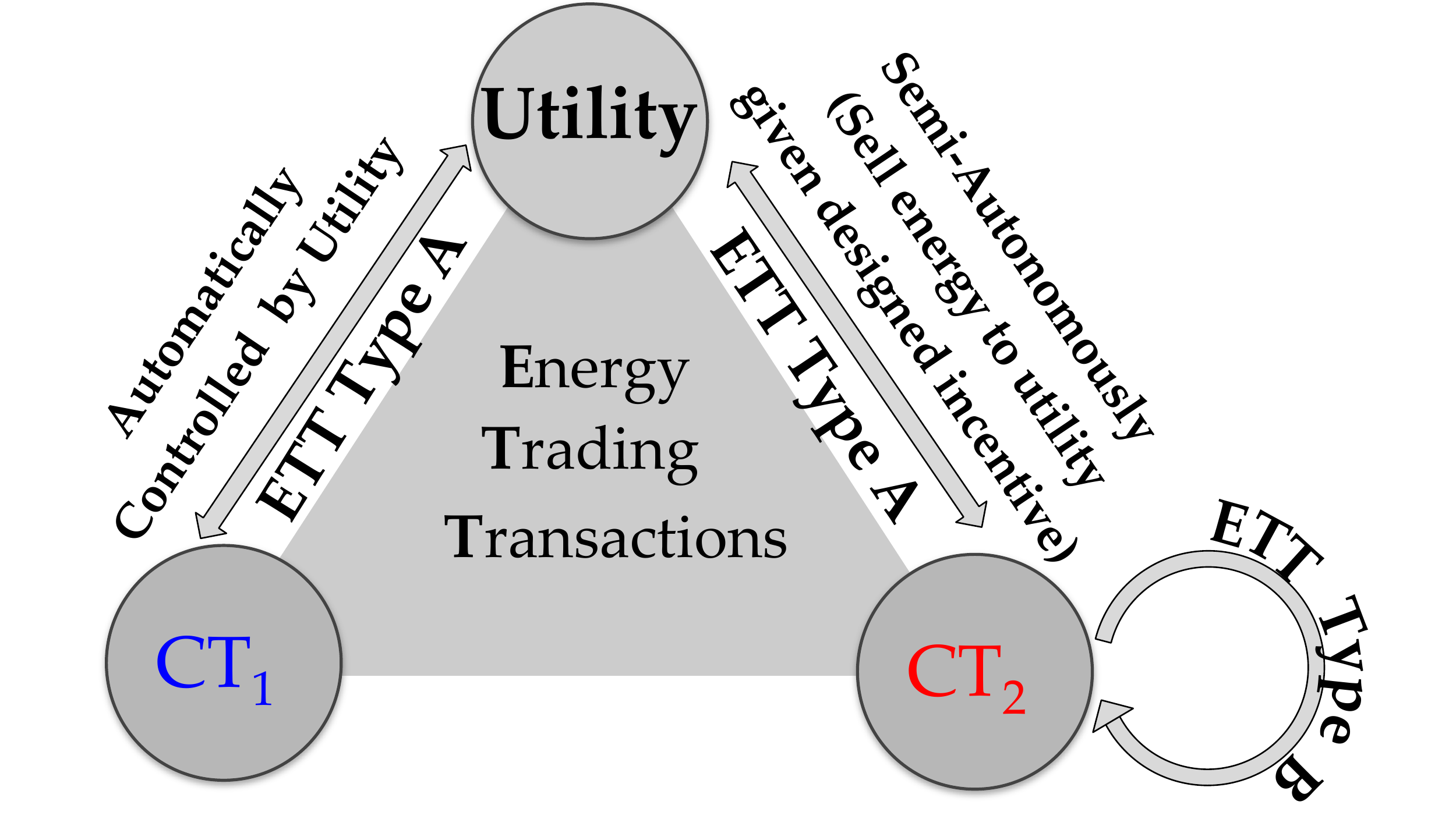}
	\caption{Types of crowdsourcees and energy trading transactions.}
	\label{fig:Energy_Trading_Transactions}
\end{figure}

The \textit{crowdsourcer}, one type of participants, here is the utility company or any other system operator, we distinguish between \textbf{two types of {crowdsourcees}} in $\mathcal{C}$. Type 1 crowdsourcees commit in the day-ahead markets (and perhaps monthly or yearly) to the crowdsourcing tasks requested by the operator. Type 1 crowdsourcees also include users who give complete control of their DERs to the operator. In return, the operator provides socio-economic incentives or discounts on the electric bill. Type 2 crowdsourcees provide near real-time adjustments or decisions based on real-time notifications and decisions from the operator. For example, the operator informs Type 2 crowdsourcees about the \textit{crowdsourced task} (e.g., charging/discharging an electric vehicle) which depends on the users' location in the network and the physical state of the grid. Type 1 crowdsourcees provide operators with day-ahead planning flexibility, in contrast with Type 2 crowdsourcees who operate on a faster timescale. The distinction between these two types of users is needed as it resembles projected market setups~\cite{wef}. We define these two types as $\mc{CT}_{1}$ and $\mc{CT}_{2}$, with $\mathcal{C} = \mc{CT}_{1} \bigcup \mc{CT}_{2}$; this is depicted in Fig.~\ref{fig:UserType}.

 We consider \textbf{two types of energy trading transactions} (ETT). \textit{Type A}: This is akin to what takes place in today's grids, where Type 1 or 2 crowdsourcees feed the grid with power. This type of transaction is solely between crowdsourcees and the network operator.  \textit{Type B}: Crowdsourcees can trade energy with each other where the seller injects power into the grid. Fig.~\ref{fig:Energy_Trading_Transactions} shows the types of crowdsourcees and transactions. Since energy production and demand response from Type 1 crowdsourcees are controlled by the operator, Type B transactions only occur among Type 2 crowdsourcees. However, Type A transactions can also take place between Type 2 crowdsourcees and the utility. The participants and the transaction types are showed in detail in Fig.~\ref{fig:Energy_Trading_Transactions}. The Brooklyn Microgrid~\cite{brooklynmicrogrid} project is an example of Type B transactions for Type 2 crowdsourcees. 

\subsection{Operational Model of Generators, Loads and DERs}~\label{sec:ModelB}
Let $i \in \mc{N}$ denote the bus index of the distribution system and $t$ denote the time-period. We consider bulk, dispatchable generation from traditional synchronous generators, renewable energy generation from solar panels, fully controllable stationary batteries, uncontrollable loads, and {shapeable} loads.

\subsubsection{Dispatchable Generators} Dispatchable generators are considered in this paper with a quadratic cost function. Dispatchable generation $S_{i,t}^g =  P_{i,t}^g + j Q_{i,t}^g$ for $i \in \mc{G}$ at $t$ are considered to have quadratic cost functions as 
$    C_{i,t}(P_{i,t}^g) = \alpha_{i,t} (P_{i,t}^g)^2 + \beta_{i,t} P_{i,t}^g + \gamma_{i,t}$
    where $\alpha_{i,t}$, $\beta_{i,t}$, and $\gamma_{i,t}$ are given parameters for the cost function of the $i$-th generator at $t$.
\subsubsection{Solar Energy Generation}

Solar panels generate real power $P_{i,t}^{r}$ for bus $i \in \mc{C}$ at $t$.
Note that $\mc{CT}_{1}$ crowdsourcees do not control whether $P_{i,t}^{r}$ is fed into the grid or not (it is controlled by the utility/operator), whereas $\mc{CT}_{2}$ crowdsourcees dictate whether to use $P_{i,t}^r$ locally or sell it to the CES operator or other users.

\subsubsection{Stationary Batteries}
    Batteries are modeled as dispatchable loads that can be controlled to withdraw or inject power. The quantity $P_{i,t}^b$ defines the output power of the batteries where $i \in \mc{C}$.  Negative $P_{i,t}^b$ implies that power is withdrawn.
    The battery operational model~\cite{munsing2017blockchains} is described as:
\begin{subequations}~\label{equ:batteries}
    \begin{align}
        E_{i,t}^{b} &= E_{i,t-1}^{b} + H_{i,t}^b  \eta_{i,in}  -  {D_{i,t}^b}/{\eta_{i,out}}         \label{particular-a}\\
        P_{i,t}^b &= D_{i,t}^b - H_{i,t}^b   \label{particular-b}\\
        0 &\leq D_{i,t}^b \leq P_{i,t,\mathrm{dis}}^b  \label{particular-c} \\
        0 &\leq H_{i,t}^b \leq P_{i,t,\mathrm{cha}}^b \label{particular-d}\\
        E^{b,\min} &\leq E_{i,t}^{b} \leq E^{b,\max}. \label{particular-e}
    \end{align}
\end{subequations}
In the above battery model, we consider a unit time-period; $\eta_{i,in}$ and $\eta_{i,out}$ represent  charging and  discharging efficiency constants. $H_{i,t}^b$ and $D_{i,t}^b$ is the charging and discharging power---both are optimization variables. The variable $E_{i,t}^{b}$, upper and lower bounded by $E^{b,\min}$ and $E^{b,\max}$, denotes the energy stored in battery at time $t$. The net power $P_{i,t}^b$  at $t$ is the difference between the power of discharging and charging. $P_{i,t,\mathrm{dis}}^b$ stands for the limitation of discharging power, $P_{i,t,\mathrm{cha}}^b$ has a similar meaning for charging power. All of variables related to batteries model are included in a single vector variable $\m {x}^b_{i,t}:=(E_{i,t}^{b},H_{i,t}^b,D_{i,t}^b,P_{i,t}^b)$.
    
\subsubsection{Uncontrollable Loads} Uncontrollable loads (lights, plug loads, street lights, et cetera) are considered to be given and are denoted by $S_{i,t}^u$ for all $i\in \mc{L}$ (loads can include reactive power), where $S_{i,t}^u = P_{i,t}^u + jQ_{i,t}^u$. 
\subsubsection{Shapeable Loads}
    We consider shapeable loads, defined by $S_{i,t}^s = P_{i,t}^s + jQ_{i,t}^s$ for $i\in  \mc{L}$, such as plug-in electric vehicles and loads from appliances with flexible power profile but fixed energy demand $E_{i,\mathrm{demand}}^s$ in 24 hours. These shapeable loads must be satisfied between $t_{i,\mathrm{start}}$ and $t_{i,\mathrm{end}}$. The model describing the shapeable loads~\cite{munsing2017blockchains} is given next.
    \begin{subequations}~\label{equ:shapeable}
        \begin{align}
            &E_{i,\mathrm{demand}}^s = \textstyle\sum_{t=1}^{T} S_{i,t}^{s} \Delta t  \\
            &S_{i,t}^{s} = 0 ,\, \text{for}\; t = 1, \ldots ,t_{i,\mathrm{start}},t_{i,\mathrm{end}}, \ldots, T\\
            &S_{i}^{s,\min} \leq S_{i,t}^{s} \leq S_{i}^{s,\max},
        \end{align}
    \end{subequations}
where $T$ is the length of the time-horizon and $\Delta t$ is the time interval. Similarly, a single vector variable $\m {x}^s_{i,t}:=(S_{i,t}^{s})$ collects variables related to shapeable loads. 
\vspace{-0.2em}
\subsection{Distribution Network Model}
For each bus $i \in \mathcal{N}$, denote $V_i = |V_i|e^{j \theta_i}$ as its complex voltage and $v_i=|V_i|^2$ as its magnitude squared. Let $s_i=p_i+jq_i$ be node $i$'s net complex power injection. Also, $p_{i}$ denotes net real power injection.  From Section~\ref{sec:ModelB}, the net real power injection for each bus $i$ at $t$ can be expressed as
    \begin{equation}~\label{equ:netpower}
    p_{i,t}  = P_{i,t}^g + P_{i,t}^b + P_{i,t}^r - P_{i,t}^u - P_{i,t}^{s}.
    \end{equation}
\noindent Similarly for the net reactive power injection. {For each line $i \in \mathcal{E} $},  we denote bus $i$'s parent and children buses as $A_i$ and $C_i$. Let $z_i = r_i + j x_i$ be its complex impedance, $I_i$ be the complex branch current from bus $i$ to $A_i$, and $l_i=|I_i|^2$ be its magnitude squared. The variable $S_i=P_i + jQ_i$ denotes the branch power flow from bus $i$ to  $A_i$. For all buses in the network, define $\m x_t:=(\m{x}^b, \m{x}^s)_t$ as a variable vector collecting the variables related to batteries and shapeable loads. Since two types of crowdsourcees are defined, $\m x_t$ is divided into two variables $\m x_{1_{t}}$ and $\m x_{2_{t}}$, which stands for the variables belong to Type 1 and Type 2 crowdsourcees and hence $\m x_t=(\m {x_1}, \m {x_2})_t$. Let $\m y_t:=(P_{i,t}^u, P_{i,t}^r)$ be a variable vector collecting the variables related to uncontrollable loads and solar energy. The preferences and setting parameters of crowdsourcees including the willingness to sell energy, constants related to batteries, solar panel or loads are communicated with the utility or the operator are denoted by $\mathcal{X}_t$. 

To model power flow in distribution networks, we use the branch flow model~\cite{farivar2013branch,low2014convex}. This model eliminates the phase angles of $V_i$ and $I_i$ and uses only $(v_i,l_i,s_i,S_i)$. 
\begin{subequations}
	\begin{align}
		v_{A_{i}}=v_{i}-2(r_{i}P_{i}+x_{i}Q_{i})+\ell_{i}(r_{i}^{2}+x_{i}^{2})\ \ \  i\in \mathcal{E} \label{subeqn:voltagebalance}\\
		\textstyle\sum_{j\in C_{i}}(P_{j}-\ell_{j}r_{j})+p_{i}=P_{i}\ \ \ i\in \mathcal{N} \label{subeqn:realpower}\\
		\textstyle\sum_{j\in C_{i}}(Q_{j}-l_{j}x_{j})+q_{i}=Q_{i}\ \ \  i\in \mathcal{E} \label{subeqn:reactpower}\\
		P_{i}^{2}+Q_{i}^{2}=v_{i}\ell_{i}\ \ \  i\in \mathcal{E} \label{subeqn:nonconvex}
	\end{align}
\end{subequations}
Due to \eqref{subeqn:nonconvex}, the branch flow model is not convex. However, the model can be convexified using the second order cone program (SOCP) relaxation~\cite{baran1989optimal} and rewritten as
\begin{equation}
\left\Vert \bmat{ 2 & 0 & 0 & 0 \\ 0 & 2 & 0 & 0\\ 0 & 0 & 1 & -1} \bmat{P_i \\ Q_i \\ v_i \\ l_i}\right\Vert \leq   \bmat{ 0 & 0 & 1 & 1} \bmat{P_i \\ Q_i \\ v_i \\ l_i} \label{equ:relaxed}
\end{equation}

The nonconvex branch flow model can be cast through convex SOCP constraints denoted by $\mathrm{CvxFlowModel}({\m z_t})$ that collects equations~\eqref{subeqn:voltagebalance}--\eqref{subeqn:reactpower} and \eqref{equ:relaxed}, and can be solved efficiently by interior-point method in polynomial time~\cite{lobo1998applications}. In this paper, all branch flow variables are collected in a single vector variable $\m {z}_t:=(\m{v},\m{l},\m{s},\m{S})_t$ at time $t$.  Tab.~\ref{table:notations} lists all variables introduced in this study. The next section introduces the CES optimal power flow formulation and incentive design.
 \begin{table}[t]
     \renewcommand{\arraystretch}{1.6}
     \fontsize{6.8}{6}\selectfont
     \caption{Notation for various DERs in CES$^*$.}
     \centering
     \begin{tabular}{ c|c }
         \textit{Symbols} & \textit{Description} \\
         \hline
        $S_{i,t}^g$ &Dispatchable generation\\
        \hline
        $P_{i,t}^{r}$ &Real power generated from solar panel\\
        \hline
        $P_{i,t}^b$ &Output power of the battery\\
        \hline
        $S_{i,t}^u$ &Apparent power of uncontrollable load\\
        \hline
        $S_{i,t}^s$ &Apparent power of shapeable load\\
        \hline
        $p_{i,t}$ &Net real power injection at each bus \\
        \hline
        $\m{x}^{b}_{i,t}$ &A variable collecting all of the variables in battery model \\
        \hline
        $\m{x}^{s}_{i,t}$ &A variable collecting all of the variables in shapeable model \\
        \hline
        $\m x_t$ &A variable collecting variables in battery and shapeable model \\
        \hline
        $\m y_t$ &A variable collecting the variables of uncontrollable loads and solar energy\\
        \hline
        $\m z_t$ &A variable collecting all of the branch flow variables \\
        \hline
        $\mathcal{X}_t$  & Preferences and setting parameters of crowdsourcees\\
        \hline    \hline
 \multicolumn{2}{l}{\footnotesize{$^*$Symbols with or without subscript ${i,t}$ have the same meaning for simplicity.}}
      \end{tabular}
     \label{table:notations}
 \end{table}

\section{CES-OPF and  Incentives Design}~\label{sec:OPF-ID}
In this section, we propose a \textbf{two-phase algorithm} minimizing the cost of generation and thermal losses by rescheduling users' shapeable loads and DERs ahead of time. The algorithm also designs localized incentives that persuade users to participate in crowdsourced energy system.  In addition, the presented algorithm supports P2P energy trading transactions between different crowdsourcees and the utility. The developed two-phase algorithm supports arbitrary P2P energy trading between prosumers and utility, resulting in a systematic way to manage distribution networks amid P2P energy trading while incentivizing crowdsourcees to contribute to this ecosystem. The algorithm also supports the operation of islanded, self-autonomous microgrids. The algorithm is described next.

The \textbf{first phase} of the algorithm is akin to day-ahead scheduling given load, solar forecasts, which belongs to \textit{optimization and grid constraints} in Section~\ref{sec:ETS}. This phase takes into account the types of crowdsourcees and their day-ahead preferences as well as the pre-scheduled ETTs among crowdsourcees. Given the day-ahead solutions from the first phase, the \textbf{second phase} reflecting \textit{market mechanism} in Section~\ref{sec:ETS} performs two significant operations. First, rectifying the mismatch in the day-ahead forecasts and hence the demand shortage/surplus by (a) obtaining more accurate, hour-ahead forecasts and (b) solving for real-time deviations in the generator and DER setpoints. Second, allowing for real-time energy transactions through the design of monetary incentives that reward crowdsourcees. Tab.~\ref{table:energyTrading} summarizes the ETT types in relevance to the two-phase algorithm. For different phases and users, the pricing mechanism also changes. Contract pricing is decided by contract between $\mc{CT}_1$ and utility, incentive pricing for $\mc{CT}_2$ is further explained in Section~\ref{sec:CES-ID}. Negotiated pricing is determined between the crowdsourcees and their neighbors.
In short, the first phase manages the larger chunk of operations, whereas the second phase deals with the mismatch in load and renewable energy generation.
The next two sections present the details of the two-phase algorithm.

 \begin{table}[t]
 	\vspace{-1em}
    \renewcommand{\arraystretch}{1.4}
    \fontsize{7.5}{6}\selectfont
     \caption{ETT types and the corresponding in relevance to the two-phase algorithm.}
     \centering
     \begin{tabular}{ c|c|c|c|c }
         & \textit{Seller} & \textit{Buyer} & \textit{Pricing Mechanism} & \textit{Optimization Phase}\\
         \hline
         {\textit{ETT Type A}}
         & $\mc{CT}_1$ & Utility & Contract pricing & Phase I\\
         \hline
         \textit{ETT Type A} & $\mc{CT}_2$& Utility & Incentive pricing & Phase II  \\
         \hline
         \textit{ETT Type B}  & $\mc{CT}_2$ & $\mc{CT}_2$& Negotiated pricing  & Phase I\\ 
         \hline        \hline
     \end{tabular}
     \label{table:energyTrading}
\end{table}
\subsection{Phase I: Day-Ahead CES Operation}~\label{sec:CES-OPF}
As discussed in Section~\ref{sec:Model}, the network operator completely controls $\mc{CT}_{1}$ users' DERs according to the signed contract, while $\mc{CT}_{2}$ users decide to participate or not in the crowdsourcing schedules based on their preferences and the offered incentives. E.g., $\mc{CT}_{2}$ users can sell their surplus solar power to the utility if designed incentive is sufficient or acceptable in the hour-ahead or real-time markets. This entails---and due to the nature of $\mc{CT}_{2}$ users---that the output from solar panels $P_{i,t}^r$, batteries $P_{i,t}^b$, and shapeable loads $P_{i,t}^s$ for users $i\in\mc{CT}_{2}$ are uncontrollable by the utility. Hence, if Type 2 crowdsourcees declare that they would not trade energy with other users (Type B transactions), then in this phase these quantities are excluded in~\eqref{equ:netpower} by setting them to zero yielding
\begin{equation}~\label{equ:netpower2}
 P_{i,t}^r = P_{i,t}^b = P_{i,t}^s = 0,i \in \mc{CT}_{2}.
\end{equation}
Otherwise, the sellers and buyers should send the energy supply-demand requests for P2P energy trading day ahead to the utility. These requests for $\mc{CT}_2$ users in time-period $t$ are expressed as constraint $\mathrm{EnergyTrading}(\m x_{2_{t}}, \m y_t)$. This constraint ultimately transforms variables $\m x_{2_{t}},\m y_t$ to mere predefined constants since the users decide to inject (or receive) a certain amount of energy into (from) the grid. 
The Crowdsourced Energy System Optimal Power Flow (CES-OPF) is formulated as
    \begin{align}
   \textbf{CES-OPF:}\;\min_{\substack{\m x_t, \m z_t\\ P^g_t}}\;&  \textstyle\sum_{t = 1}^{T} 
    J_t(\m x_{t},\m z_t, \m P^g_t) \notag \\
     \mathrm{s.t.}\; &\eqref{equ:batteries}-\eqref{equ:netpower}, \eqref{equ:netpower2},\m y_t = \m y_t^{\mathrm{f-24hr}}, \m x_t \in \mathcal{X}_t~\label{equ:CES-OPF} \\
     &\mathrm{CvxFlowModel}({\m z_t}),{ \m z^{\min}_{t}} \leq {\m z_t} \leq {\m z^{\max}_{t}}\notag    \\
     &P^g_t \in \mc{P}, \; \mathrm{EnergyTrading}(\m x_{2_{t}},\m y_t) \notag.
     \end{align}
The objective function of CES-OPF at time $t$ is defined as $$  J_t(\m x_{t}, \m z_t, \m P^g_t) = \sum_{i = 1}^{n_g} C_{i,t}(P_{i,t}^g) + \sum_{i = 1}^{|\mathcal{E}|} l_{i,t} r_i + \sum_{i =1}^{|\mathcal{CT}_1|} U_i(x_{t}).$$ 
 The objective is to minimize the generator's cost function, given by  $\sum_{i = 1}^{n_g} C_{i,t}(P_{i,t}^g)$, in addition to the thermal losses that are characterized by $ \sum_{i = 1}^{|\mathcal{E}|} l_{i,t} r_i$, and crowdsourcees' disutility function  $U_i(x_{t}) =  u_i  (S_{i,t}^{s} -  S_{i}^{s,\max})^2, \forall t \leq T_{\mathrm{set}}$ designed to compensate for the inconvenience caused by rescheduling shapeable load. The  parameter $u_i \in [0,1]$ stands for the urgency to finish a certain task before a setting time $T_{\mathrm{set}}$; the $S_{i}^{s,\max}$ is the same parameter appearing in~\eqref{equ:shapeable}; and $u_i$ is parameter determined by users through preferences  $\mathcal{X}_t$. 
 
 The CES-OPF captures the cost of power losses between two peers through the second term of $J_t(\cdot)$ which sums the losses for all lines $\mathcal{E}$ in a distribution network. These lines include  the distribution lines between any two users/peers, including traditional energy consumers. Preferences set by users are included in $\mathcal{X}_t$ and are assumed to be linear and time-dependent; $\m y_t^{\mathrm{f-24hr}}$ is the day-ahead uncontrollable load and solar energy forecasts. Constants $\m z^{\min}_t$ and $\m z^{\max}_t$ are lower and upper bounds on branch flow model variable ${\m z_t}$; i.e. , the voltage in p.u. at each node is in $[0.95\ 1.05]$. The linear ramp constraints and upper/lower bounds on $\m P^g_t$ are denoted by $\mc{P}$.

The CES-OPF can be decomposed into small optimization sub-problems by decoupling variables and constraints---the overall problem can be then solved through a decentralized alternating direction method of multipliers (ADMM) algorithm; see~\cite{peng2014distributed}. Another approach is to simply solve CES-OPF in a centralized fashion after requesting the user's preferences $\mathcal{X}_t$ ahead of time for medium- or small-scale distribution networks and microgrids. Another way of making CES-OPF more computationally tractable is to replace the convexified branch flow model with the $\mathrm{LinDistFlow}(\m z_t)$ model~\cite{turitsyn2011options} which is linear in $\m z_t$; this transforms CES-OPF to a quadratic program that can be solved for large-scale networks. 

After solving CES-OPF, we obtain the equilibrium $S_{i,t}^{\mathrm{g,eq}} = P_{i,t}^{\mathrm{g,eq}} + j Q_{i,t}^{\mathrm{g,eq}}$ and $\m x^{\mathrm{eq}}_{1_{t}}$ which includes $P_{i,t}^{\mathrm{b,eq}}$ and $S_{i,t}^{\mathrm{s,eq}}$. This entails that the utility-scale generation, batteries and shapeable loads belonging to $\mc{CT}_{1}$ users will be fixed with this equilibrium for the next 24 hours. To compensate crowdsourcees for their contributions,  
the distributed locational marginal price (DLMP)---the time-varying electricity price for users at various buses in the network---is computed by finding the dual variables associated with the real power balance constraint in the convexified branch flow model, and denoted by $\lambda_{i,t}^{\mathrm{eq}}$.
\subsection{Phase II: Real-Time CES Incentives Design}~\label{sec:CES-ID}
As outlined in Section~\ref{sec:CES-OPF}, we solve CES-OPF and obtain setpoints for utility-scale power plants and Type 1 crowdsourcees, knowing that some enery trading transactions will take place between crowdsourcees. In this section, the presented crowdsourcing incentive design performs the two key functions: (a) Incentivizes Type 2 crowdsourcees to sell excess solar power to the utility; (b) Mitigates and balances the unexpected load and solar output fluctuations due to the forecast error in the grid. The formulation presented in this section is solved every hour or less, depending on the availability of hour-ahead forecasts and the operator's preference.

Here, we outline the design of crowdsourcing incentives that provide near real-time ancillary services to relieve real-time demand shortage or surplus---and hence the additional incentives which based on the amount of energy provided to the grid are offered for $\mc{CT}_{2}$. For $i \in \mc{CT}_2$, the amount of energy provided to the grid is depicted by the net injection power $P_{i,t}^{\mathrm{ni}}$ and computed as 
 \begin{equation}~\label{equ:netpower3}
 P_{i,t}^{\mathrm{ni}} = P_{i,t}^{r} - P_{i,t}^{s} + P_{i,t}^{b}, \;\;\;\;i \in \mc{CT}_{2}.
 \end{equation}
This indicates when solar panels produce more power, and the shapeable load reduces, more net injected power can be sold to the utility or other crowdsourcees through energy trading. Here, for $ i \in \mc{CT}_{2}$, shapeable loads and batteries cannot be scheduled 24 hours ahead since no contract exists between Type 2 crowdsourcees and the utility. 
Hence, $P_{i,t}^{s}$ and $P_{i,t}^{b}$  belonging to variable $\m x_{2_{t}}$ are treated now as uncontrollable loads for $\mc{CT}_2$ in Phase II. In addition, solar energy is also known ahead of time. Hence, $P_{i,t}^{\mathrm{ni}}$ is known and not an optimization variable for Type 2 crowdsourcees from (\ref{equ:netpower3}). The crowdsourcing incentive design routine for crowdsourcees $i$ at time $t$ is formulated as
\begin{align}
\textbf{CES-ID:}\;\min_{\substack{\m x_t, \m z_t \\  \m P^g_t , \m \lambda_t^a\\ \m b_{t} }}\;&\sum_{i = 1}^{n_g} C_{i,t}(P_{i,t}^g -P_{i,t}^{\mathrm{g,eq}}) + \sum_{i = 1}^{|\mathcal{E}|} l_{i,t} r_i + \sum_{i = 1}^{|\mc{CT}_{2}|} b_{i,t} \notag \\
\mathrm{s.t.}\;&\eqref{equ:batteries}-\eqref{equ:netpower},\eqref{equ:netpower3},\m x_{1_{t}} = \m x^{\mathrm{eq}}_{1_{t}}, \m x_{2_{t}} \in \mathcal{X}_{2_{t}}\notag    \\
&\m y_t = \m y_t^{\mathrm{f-1hr}},{ \m z^{\min}_{t}} \leq {\m z_t} \leq {\m z^{\max}_{t}}\label{equ:CES-ID}\\
& \mathrm{CvxFlowModel}({\m z_t}),P^g_t \in \mc{P} \notag\\
& b_{i,t} = P_{i,t}^{\mathrm{ni}}(\lambda_{i,t}^{\mathrm{eq}} + \lambda_{i,t}^a),b_{i,t} \geq 0, i \in \mc{CT}_{2}  \notag \\
&\textstyle\sum_{i = 1}^{|\mc{CT}_{2}|} b_{i,t} \geq b_{t}^{\mathrm{total}},i \in \mc{CT}_{2}. \notag
\end{align}
In CES-ID, the objective is to minimize (a) the deviation in the cost of generation from the day-ahead operating point, (b) the network's thermal losses, and (c) the budget $\sum_{i = 1}^{|\mc{CT}_{2}|} b_{i,t}$ (in \$) which the operator has allocated to spend on the real-time incentives at the feeder level. 
The constraints are explained as follows. We set variables $P_{i,t}^b, S_{i,t}^s \in \m x_{1_{t}}$ to the equilibrium  $P_{i,t}^{\mathrm{b,eq}},S_{i,t}^{\mathrm{s,eq}} \in \m x^{\mathrm{eq}}_{1_{t}}$ which is obtained by CES-OPF to schedule DERs that are controlled by the utility. For $i \in \mc{CT}_2$, the willingness to sell energy to the utility is set in preference $\mathcal{X}_{2_{t}}$ which sent to system operator. 
 The constraints on $\m y_t,\m z_t, P^g_t$ are the same as CES-OPF~\eqref{equ:CES-OPF} except that $\m y_t$ is set to the hour-ahead (or shorter) available forecast $\m y_t^{\mathrm{f-1hr}}$.

Besides the optimization variables mentioned above,  we consider that Type 2 crowdsourcees receive the final incentive price $\lambda_{i,t}^{\mathrm{eq}} + \lambda_{i,t}^a$ where $\lambda_{i,t}^a$,  additional variable, is an adjustment price which varies with the net energy injected to grid and location of $\mc{CT}_2$; $\lambda_{i,t}^{\mathrm{eq}}$ is DLMP computed by CES-OPF. 
The variable budget $b_{i,t}$ for $i \in \mc{CT}_2$ at $t$ is equal to $P_{i,t}^{\mathrm{ni}}(\lambda_{i,t}^{\mathrm{eq}} + \lambda_{i,t}^a)$, which is always greater than 0. As mentioned $P_{i,t}^{\mathrm{ni}}$, $\lambda_{i,t}^{\mathrm{eq}}$ are known. When the crowdsourcee $i$ has no energy to sell to utility ($P_{i,t}^{\mathrm{ni}} \leq 0$), variable $\lambda_{i,t}^a$ is forced to approach $-\lambda_{i,t}^{\mathrm{eq}}$ to make $b_{i,t}$ as $0^{+}$ (a small positive value which is approximately close to zero). Hence no incentive is offered to those who inject no power into the grid. When $P_{i,t}^{\mathrm{ni}} > 0$ which means crowdsourcee $i$ at $t$ has excess energy to sell, variable $\lambda_{i,t}^a$ is forced to be small while also minimizing the final incentive price $\lambda_{i,t}^{\mathrm{eq}} + \lambda_{i,t}^a$ and budget $b_{i,t}$ for all Type 2 crowdsourcees. 
At time $t$, the total budget for $\mc{CT}_2$ is $b^{\mathrm{total}}_t$, which can be set as a reasonable value. For example, this can be set to the cost for dispatchable generation to produce $\sum_{i = 1}^{|\mc{CT}_{2}|} P^{\mathrm{ni}}_{i,t}$. Further explanations and examples are presented in Section~\ref{sec:Results_Discussions}.

Notice that both CES-OPF and CES-ID are based on branch flow model which is convex, and can be solved with great efficiency in polynomial time by interior-point optimizer. The CES-ID is solved hourly, and the computed incentives are sent to users at the end of the day. Thus, the energy trading (Type A transactions) between $\mc{CT}_2$ users and the utility is finished. The transactions are done by the assist of blockchain, which is explained in next section.
\section{Blockchain and Smart Contracts implementation for CESs}~\label{sec:BlockChain}
In this section, we discuss an implementation for blockchain that is scalable to accommodate millions of crowdsourcees and energy trading transactions. An algorithm to integrate the optimization models in Section~\ref{sec:OPF-ID} with this blockchain implementation is also presented.
\subsection{Blockchain and Smart Contracts Implementation for CESs}
While Tab.~\ref{table:BlockchainType} summarizes the attributes of different blockchain platforms, this section identifies the properties most applicable for the proposed crowdsourced energy system model and algorithms introduced in Section~\ref{sec:Model} and \ref{sec:OPF-ID}. Specifically, the blockchain platform must adequately address the goals to incorporate a precise set of CES users, the computational requirements of the crowdsourced energy system algorithms, the performance of the consensus algorithms, and the privacy demands of the users. The CES requirements and blockchain properties for each of these domains are identified in Tab.~\ref{tab:newBC}.
\begin{table}[t]
  \centering
      \renewcommand{\arraystretch}{1.1}
  \caption{CES requirements mapping to blockchain features.}
  	\begin{tabular}{m{4.5em}|m{11em}|m{11em}}
  	{} & {\textbf{CES Requirements}} & {\textbf{Blockchain Features}} \\
  	\hline
  	\hspace{-0.8em}{\textbf{Participants}} & The CES will be operated for a distribution grid, so users will be confined to a geographic area users & Permissioned chain as users should be restricted to those currently within that distribution area \\
  	\hline
  	\hspace{-1.2em}\textbf{{Computation}} & CES must require performing non-linear optimizations such as solving power flow and economic dispatch & Efficient smart contracts requiring the ability to execute Turing complete programs on large quantities of data without heavy cost \\
  	\hline
  	\hspace{-0.8em}\textbf{Consensus} & Minimal energy usage to ensure energy sustainability goals of CES & Avoid computationally expensive PoW consensus algorithms \\
  	  	\hline
  	\hspace{-0.8em}\textbf{Privacy} & Crowdsourcee preferences and usages likely exposes privacy data & Permissioned model that protects crowdsourcee data from external observers \\
  	  	\hline   	  	\hline
  \end{tabular}%
  \label{tab:newBC}%
\end{table}%
\begin{figure*}[t]
	\centering
	\includegraphics[scale=0.425]{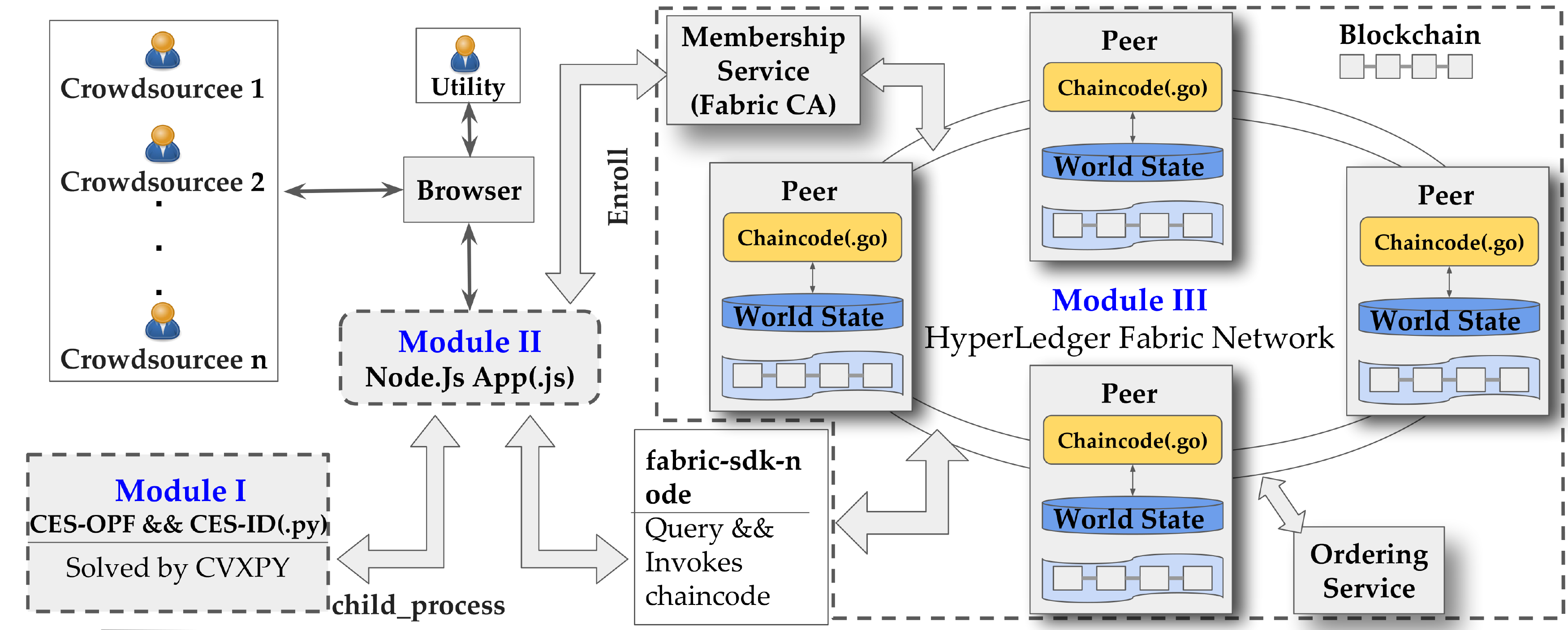}
	\caption{Architecture of combining blockchain and smart contract with the optimization formulations presented in this paper.}
	\vspace{-0.5cm}
	\label{fig:blockchain}
\end{figure*}
Based on this analysis, the Hyperledger is selected to meet the required CES requirements and necessary blockchain features. As previously mentioned, Hyperledger uses RBFT for consensus, which should minimize the energy required for each transaction. Furthermore, Hyperledger's permissioned model ensures that the participants are restricted to those within the distribution grid’s service region, and also prevents the exposure of privacy data from crowdsourcees. Finally, the smart contracts can be implemented through the \texttt{chaincode} mechanisms, which does not require the per-operation execution costs that are enforced on other public blockchains. 

This, unlike other blockchain applications, still requires a central authority---the utility company or the system operator to manage the grid, provide technical supports for each small-scale energy trading, clear the market, and ensure there is no violation of any technical constraints (e.g., distribution line limits). Small-scale energy trading without a central authority can take place (see~\cite{Cottrell_2017}), yet the scaling of these transactions to include thousands of people and millions of daily energy transactions without the utility coordinating the communication among small-scale energy trading systems is remote in today’s markets. To this end, the presented architecture in this paper requires a central authority to manage the grid but can also autonomously be run in islanded microgrids as we showcase in the case studies section.

\subsection{Blockchain Implementation using Hyperledger Fabric} \label{sec:Implementation}
We integrate and implement blockchain and smart contracts with the optimization models given in Section~\ref{sec:OPF-ID}. This is shown in Fig.~\ref{fig:blockchain}. The presented CES implementation consists of three modules---surrounded by the dotted lines in Fig.~\ref{fig:blockchain}. Module I, corresponding to \textit{optimization and grid constraints} in~Section~\ref{sec:ETS}, includes the optimization problems in Section~\ref{sec:OPF-ID} which are coded by \texttt{CVXPY}~\cite{CVXPY}. Module II is a \texttt{Node.js} application, also take care of the communication between Python-written optimization problem and Module III. This process is finished by the \texttt{child\_process} standard library which generates a python process and computes the solutions to CES-OPF~\eqref{equ:CES-OPF}, CES-ID~\eqref{equ:CES-ID} while returning results back to \texttt{Node.js} program.

Module III, the \textit{information system} in Section~\ref{sec:ETS}, is implemented by the IBM Hyperledger Fabric Network deployed in cloud to provide the blockchain service. The network consists of many peers that communicate with each other, runs smart contracts called \texttt{chaincode} which is written by Go language, holds state and ledger data. Peers in the Hyperledger Fabric Network are different from the ones in the other blockchain implementations. The roles of peers relate to the life-cycle of transactions which is one key difference between Hyperledger Fabric and many other blockchain platforms. The life-cycle of a transaction in other blockchain platforms is usually {order-execute}, which means that transactions are added to the ledger in a specific order and executed sequentially. But in Hyperledger Fabric, it is a three-step process: {execute-order-validate}. First, transactions are executed in parallel considering any order. Second, they are ordered by an ordering service. Third, each peer validates and applies the transactions in sequence. The roles of peers also have a strong relationship with robust privacy and permission support, the reader is referred to~\cite{fabricpeers} for more information.

The crowdsourcees shown in Fig.~\ref{fig:blockchain} are the end-users in the distribution network and can perform energy trading. Thousands of crowdsourcees are allowed to connect and sign up to the Fabric network via a browser after receiving a code from the operator. The operator also can log in via browser to manage the overall system---screen shots are given in the next section showing the graphical user interface. After enrolling in the network via \texttt{Fabric-CA}~\cite{fabric-ca}, a certificate needed for enrollment through a software development kit (SDK), crowdsourcees can communicate with the network through \texttt{fabric-sdk-node}~\cite{fabric-sdk-node}, update their preferences to blockchain and store it in World State~\cite{worldstate} which is the database. Peers in Hyperledger are used to commit transactions, maintain the world state and a copy of the ledger (consists of blocks). The \texttt{chaincode} in Hyperledger Fabric is deployed into peers and is executed as a user satisfies their commitments. Then, \textit{ordering service}, akin to mining in Bitcoin, generates new blocks in Fabric. Every peer updates their local blockchain after receiving ordered state updates in the form of blocks from the ordering service. In this way, the order and number of blocks, a form of blockchain, are maintained and synchronized for all peers. The ETTs records are included in blockchain stored  at each peer's repository and protected by this mechanism.

This specific implementation is endowed with the following characteristics: \textit{(i)} Scalable to million of crowdsourcees, \textit{(ii)} Requires little understanding of the blockchain technology from the users' side, \textit{(iii)} Communicates seamlessly with any optimization-based formulation, and \textit{(iv)} Requires very little energy to run blockchain.  
Algorithm~\ref{cecalgo} illustrates how the developed optimization routines are implemented with blockchain and smart contracts. 
\begin{algorithm}[t]
	\small    \caption{Blockchain-Assisted CES Operation}\label{cecalgo}
	\begin{algorithmic}
		\STATE \textbf{Phase I: } 
		\STATE Obtain crowdsourcees preferences $\mc{X}_t$
		\STATE Request/obtain day-ahead P2P ETT requests via blockchain implementation developed (Fig.~\ref{fig:blockchain})
		\STATE Estimate day-ahead forecasts $\m y_t^{f-24hr}$
		\STATE Solve CES-OPF~\eqref{equ:CES-OPF} and obtain generator and DER schedules
		\STATE  Establish Type A ETTs smart contracts for users $i \in \mc{G} \bigcup \mc{CT}_{1}$
		\STATE  Establish Type B ETTs smart contracts for users $i \in \mc{CT}_{2}$
		\STATE \textbf{Phase II: } 
		\WHILE {$ t \in 1,\ldots,24$ hrs}
		\STATE  Select Type 2 crowdsourcees willing to sell solar power to the utility at time $t$ according to the preferences $\mathcal{X}_{2_t}$
		\STATE  Obtain hour-ahead forecasts $\m y_t^{f-1hr}$
		\STATE  Solve CES-ID~\eqref{equ:CES-ID} at time $t$ 
		\STATE Communicate to crowdsourcees $i\in \mc{CT}_{2}$ incentives $\lambda_{i,t}^{\mathrm{eq}} + \lambda_{i,t}^{a}$
		\STATE  Establish Type A ETTs smart contracts for users $i \in  \mc{CT}_{2}$
		\ENDWHILE
		\STATE Reconcile payments weekly or monthly
	\end{algorithmic}
\end{algorithm}
\section{Case Studies}\label{sec:Case-Studies}
\subsection{Simulation Setup}\label{sec:Setup}
\begin{figure}[t]
	\centering
	\includegraphics[scale=0.25]{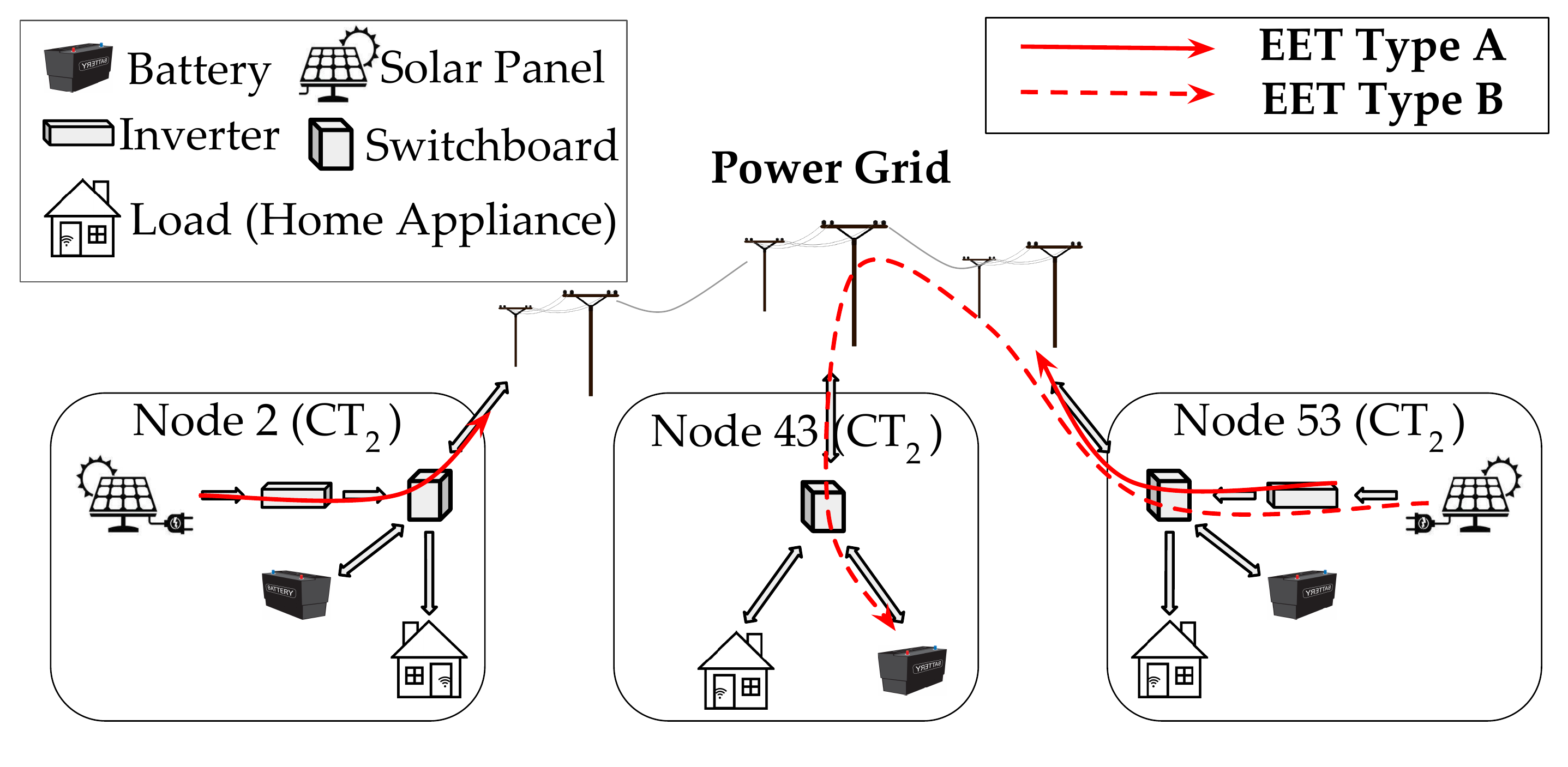}
	\caption{Scenarios of energy trading transactions.}
	\label{fig:connection}
\end{figure}

The numerical tests are simulated in Ubuntu 16.04.4 LTS with an Intel(R) Xeon(R) CPU E5-1620 v3 @ 3.50 GHz. We use the Southern California Edison (SCE) 56-bus test feeder~\cite{peng2013optimal} as a distribution network. Reasonable uncontrollable load profile $\m P^u$ is generated for $T=24\;\mathrm{hrs}$ from California Independent System Operator (CAISO)~\cite{CAISO} and normalized to ensure that the optimization problems have feasible sets for different time-periods. We modify SCE 56-bus test feeder as shown in Fig.~\ref{fig:UserType} and place stationary batteries, solar panels, uncontrollable and shapeable loads at each bus in the network; see Fig.~\ref{fig:connection}. Similar to~\cite{munsing2017blockchains}, batteries are set up with a power capacity of 80\% of the peak uncontrollable load at the bus, an 4-hour energy storage capacity with 20\% initial energy storage. We assume that the solar generation power profile is given and contributes to 50\% of the uncontrollable load at peak for each bus. Shapeable loads have net energy demand that is up to 20\% the peak power consumption of the uncontrollable loads and can be charged for 4--8 hours. The scheduling time of shapeable loads is from 8 am to 11 pm. 

We also assume that each bus is connected to a crowdsourcee of Type 1 ($\mc{CT}_1$) or Type 2 ($\mc{CT}_2$). We make the following assignment: If the number of a bus is a prime number, then the user belongs to $\mc{CT}_2$, otherwise they belong to $\mc{CT}_1$ (we have $|\mc{CT}_1| = 40$ and $|\mc{CT}_2| = 16$). From the above setup, Nodes 2, 43 and 53 belong to $\mc{CT}_2$ in Fig.~\ref{fig:connection}. 
As we present in Tab.~\ref{table:energyTrading}, two types of energy trading transactions  take place in crowdsourced energy systems. Type A transactions occur between $\mc{CT}_1$ or $\mc{CT}_2$ with utilities, while the trading transactions among $\mc{CT}_2$ users are Type B transactions. In Fig.~\ref{fig:connection}, we present two scenarios of energy trading transaction for further explanation: \textit{(i)} ETT Type A where Node 2 decides to sell excess solar energy to the utility, \textit{(ii)} ETT Type B where Node 43 chooses to buy energy from Node 53. The next section presents the outcome of the two-phase optimization discussed in Section~\ref{sec:OPF-ID}.
\begin{figure}[t]
    \centering 
    \includegraphics[scale=0.33]{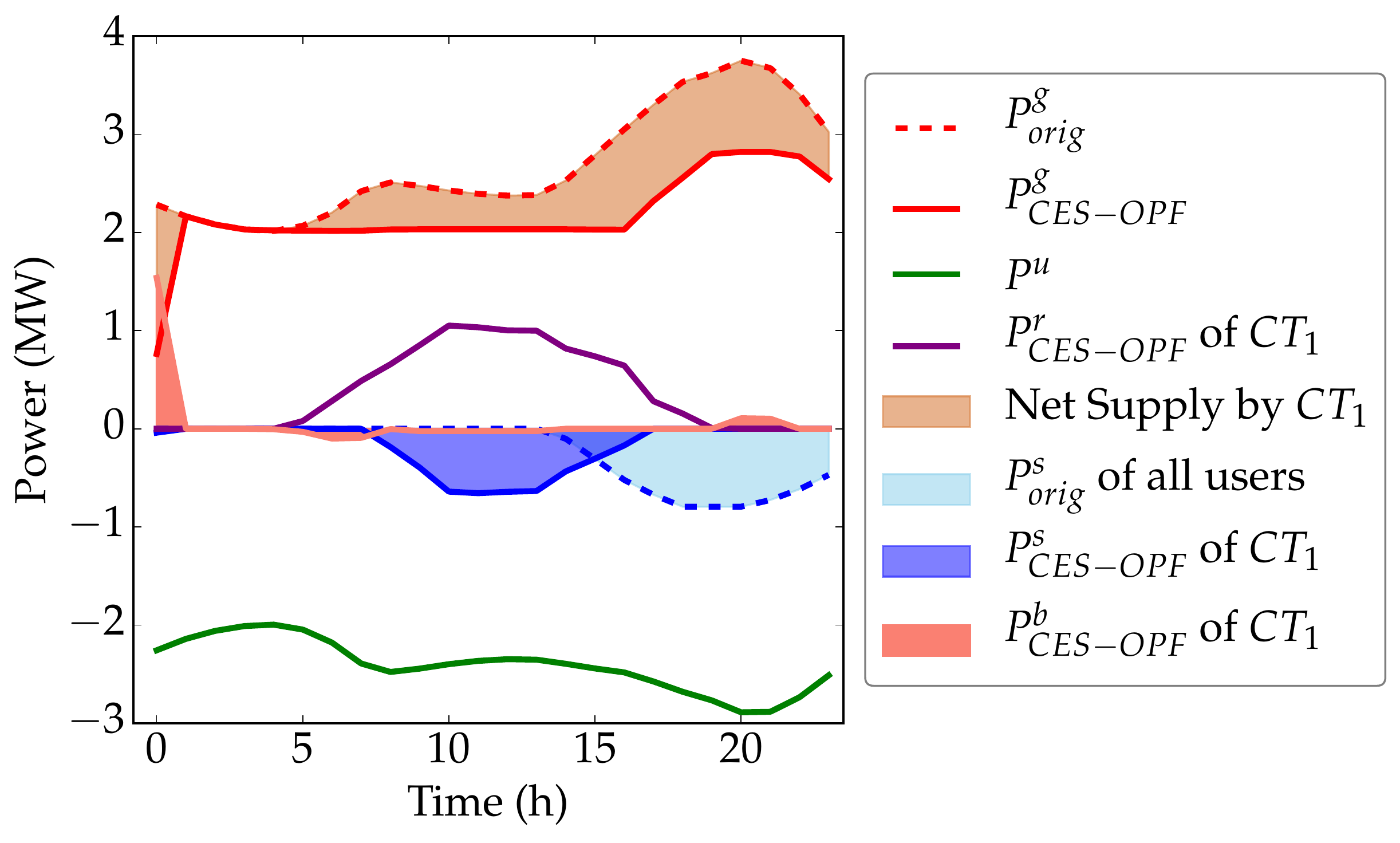}
    \caption{Aggregate load profile and generation after solving CES-OPF~\eqref{equ:CES-OPF}. }
    \label{fig:sumofnetwork_CESOPF}
  \vspace{-0.4cm}
\end{figure}
\begin{figure}[t]
    \centering
    \includegraphics[scale=0.32]{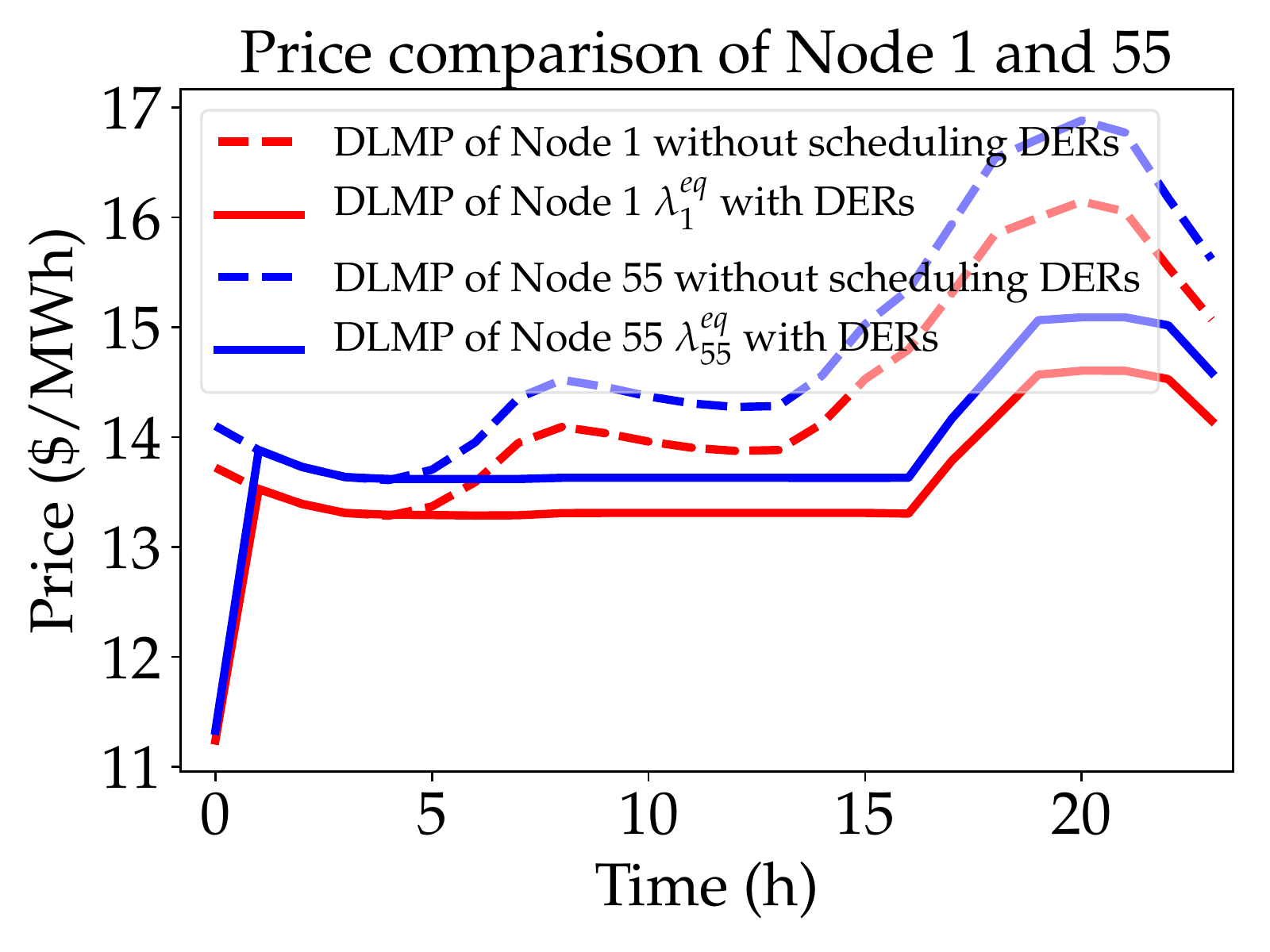}
    \caption{Price comparison of Node 1 and 55 before and after CES-OPF.}
    \label{fig:Price_comparison}
\end{figure}
\subsection{Results and Discussions}~\label{sec:Results_Discussions}
In order to present the effectiveness of our algorithm, we compare the cases with and without considering the energy trading among crowdsourcees based on the modified SCE 56 bus test feeder illustrated in Section~\ref{sec:Setup}.
\subsubsection{Phase I: Day-Ahead CES Operation}
solving CES-OPF~\eqref{equ:CES-OPF}. Fig.~\ref{fig:sumofnetwork_CESOPF} shows $P^u$, $P^s_\mathrm{orig}$ and $P^g_\mathrm{orig}$ (the aggregate uncontrollable load, shapeable load, and the output of generator) when our algorithm is not applied---in the absence of energy crowdsourcing or trading between crowdsourcing. Fig.~\ref{fig:sumofnetwork_CESOPF} also shows the aggregate load profile and generation after solving the CES-OPF for $T=24\;\mathrm{hrs}$. The figure shows that battery variable $P^b_\mathrm{CES-OPF}$ charges when the solar panel produces and injects power $P^r_\mathrm{CES-OPF}$ into network. The reason why the curve of $P^b_\mathrm{CES-OPF}$ does not change significantly is that the solar panels do not generate enough energy in this setup. Hence the algorithm is less inclined to store energy into batteries. As for the scenarios when the solar panel produces enough energy, please refer to { Fig.~\ref{fig:island} in the section of Islanded Microgrid Test (\ref{sec:islandedtest}).  Fig.~\ref{fig:sumofnetwork_CESOPF} indicates that shapeable loads of $\mc{CT}_1$ are rescheduled to $P^s_\mathrm{CES-OPF}$. The updated power generation $P^g_\mathrm{CES-OPF}$ is smaller than $P^g_\mathrm{orig}$ due to the injections of solar power, scheduling of batteries and shapeable loads from crowdsourcees $\mc {CT}_1$.
Fig.~\ref{fig:Price_comparison} presents the changes in the  DMLPs with and without scheduling DERs in the distribution network through CES-OPF~\eqref{equ:CES-OPF} for Nodes 1 and 55. The DLMPs for both nodes are smaller due to the net injection from Type 1 crowdsourcees (shaded orange area in Fig.~\ref{fig:sumofnetwork_CESOPF}). This illustrates how the DLMP price becomes lower when rescheduling DERs and injecting renewable energy into the grid.
 \begin{figure}[t]
     \centering \includegraphics[scale = 0.35]{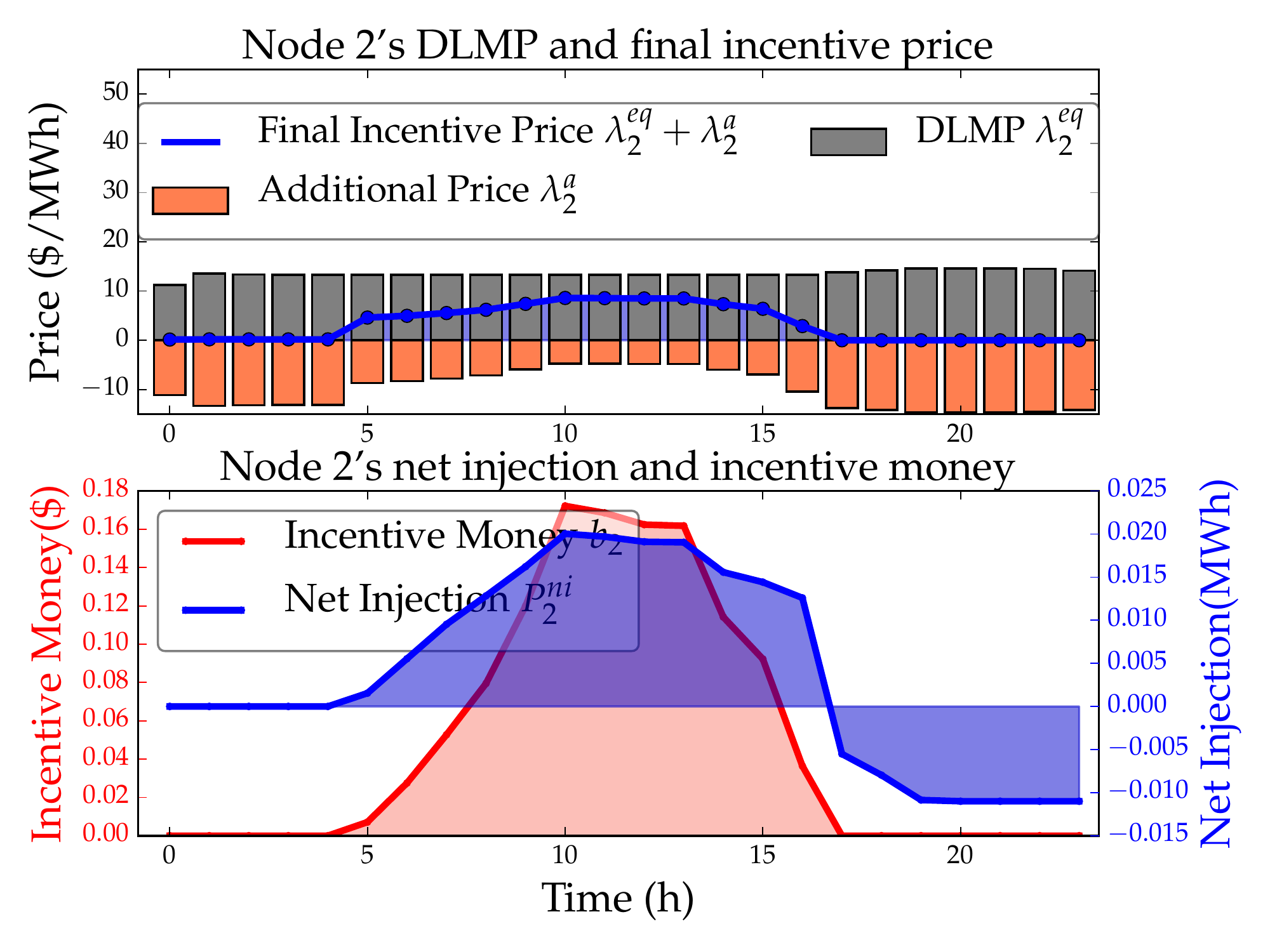}
     \caption{Final incentive price, net injection and incentive money for Node 2.}
     \label{fig:Node2_incentive_all}
 \end{figure}
 \begin{figure}[t]
     \centering
	\vspace{-0.4cm}
         \includegraphics[scale = 0.35]{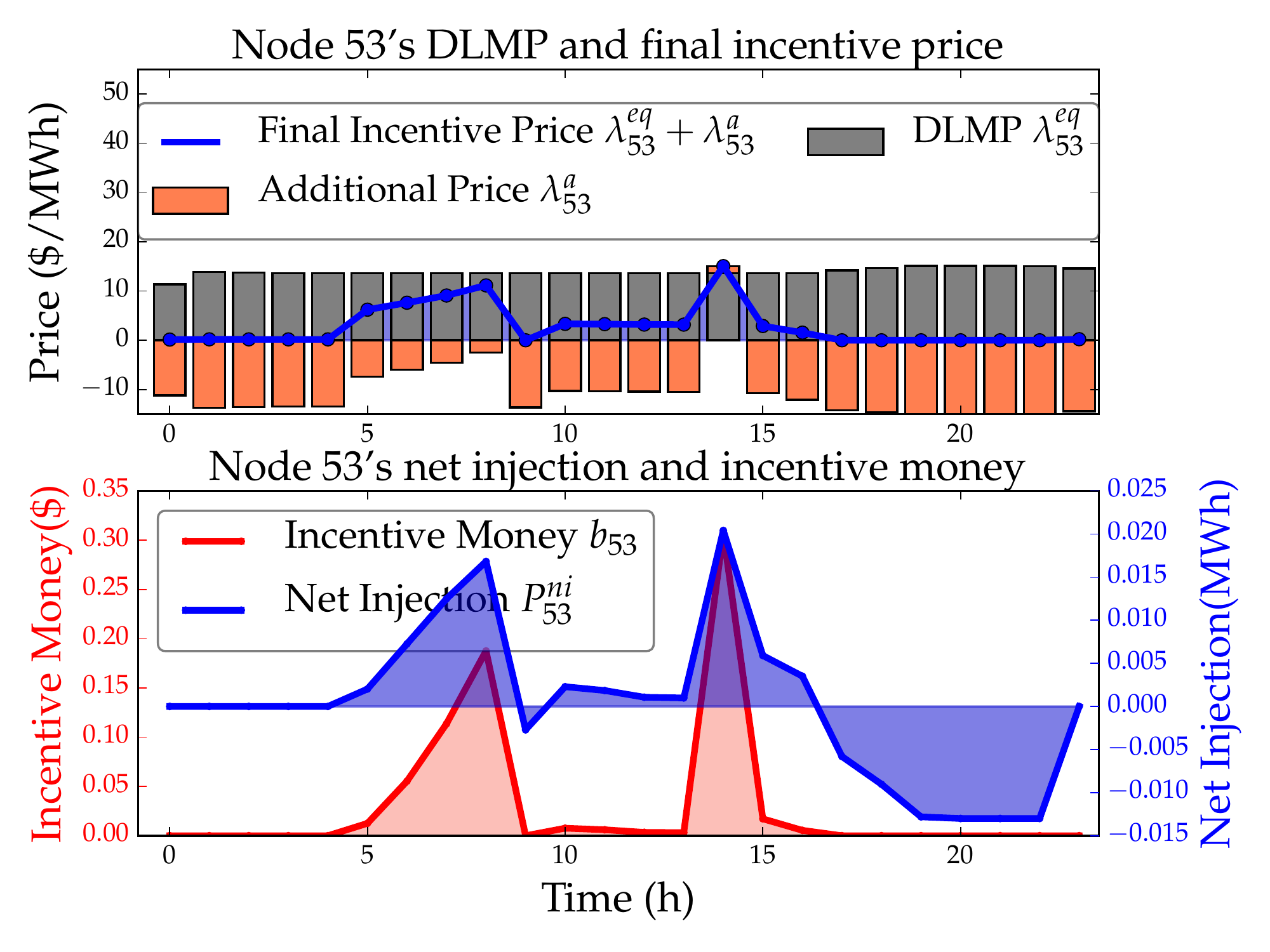}
     \caption{Final incentive price, net injection and incentive money for Node 53.}
     \vspace{-0.5cm}
     \label{fig:Node43_53_incentive_all}
 \end{figure}
\subsubsection{Phase II: Real-Time CES Incentives Design}
CES-ID is solved once every hour but it can also be solved every 5--15 minutes depending on the availability of accurate weather/load forecasts. The monetary rewards offered to Type 2 crowdsourcees are obtained from CES-ID. We assume that the crowdsourcees of Type 2 at Nodes 2, 43, and 53 accept the designed incentives. 

Fig.~\ref{fig:Node2_incentive_all} shows the final incentive price, net injection, and overall incentive money for Node 2. The time-varying nature of the final incentive price of a node is due to variations of its DLMP and its net injection. We assume that the solar panel produces energy between 6 am and 7 pm. The solar panel of Node 2 produces solar power and incentives are earned by the customer between 6 am and 2 pm as shown in Fig.~\ref{fig:Node2_incentive_all}. However, the load at Node 2 starts to consume energy at 5~pm making the net injection of Node 2 is 0 MWh. Hence, no monetary incentives are offered from 7 pm to 11 pm. Fig.~\ref{fig:Node43_53_incentive_all} presents the results for Type B transactions for $\mc{CT}_2$ user at Node 53. The user at Node 43 decides to charge the battery at a  constant charging rate between 9 am and 2 pm, and the excess solar energy produced from Node 53's solar power can satisfy this demand shortage. Notice that Node 43 only consumes energy while Node 53 earns incentive rewards from the utility and negotiated money from Node 43 during different time periods. The transaction details between these crowdsourcees are summarized in Tab.~\ref{table:node53energytrading}.

    \begin{table}[t]
    \renewcommand{\arraystretch}{1.4}
    \fontsize{7.5}{6}\selectfont
    \caption{Transaction details for Node 53.}
    \centering
    \begin{tabular}{ c|c|c|c|c|c }
        \textit{Time} & \textit{Seller} & \textit{Buyer} & \textit{Energy} & \textit{ETT Type} & \textit{Phase}\\
        \hline
        {{6 am--9 am}}
        & Node 53 & Utility & 0.0385 MWh &  Type A & Phase II\\
        \hline
        {9 am--2 pm} & Node 53 & Node 43 & 0.119 MWh & Type B & Phase I\\
        \hline
        {14 pm--5 pm}  & Node 53 & Utility & 0.062 MWh   &  Type A & Phase II\\
        \hline
        \hline
    \end{tabular}
    \label{table:node53energytrading}
\end{table}

\begin{figure}[t]
    \vspace{-0.2cm}
    \centering \includegraphics[scale = 0.30]{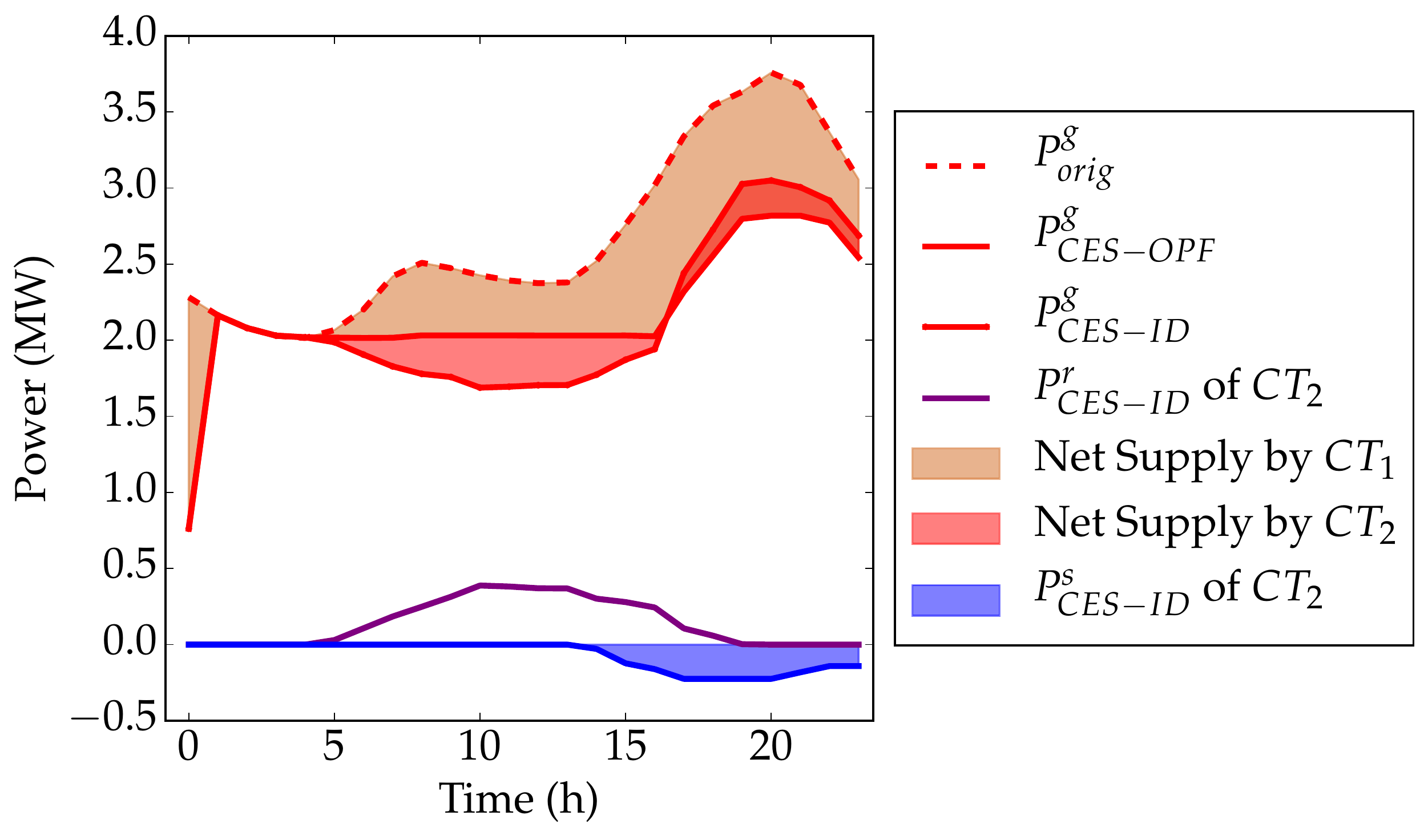}
    \caption{Aggregate load profiles and generation after Algorithm~\ref{cecalgo} terminates and the incentives are designed.}
    \label{fig:sumofnetwork_CESID}
    \vspace{-0.5cm}
\end{figure}
  \begin{figure}[t]
	\centering \includegraphics[scale = 0.30]{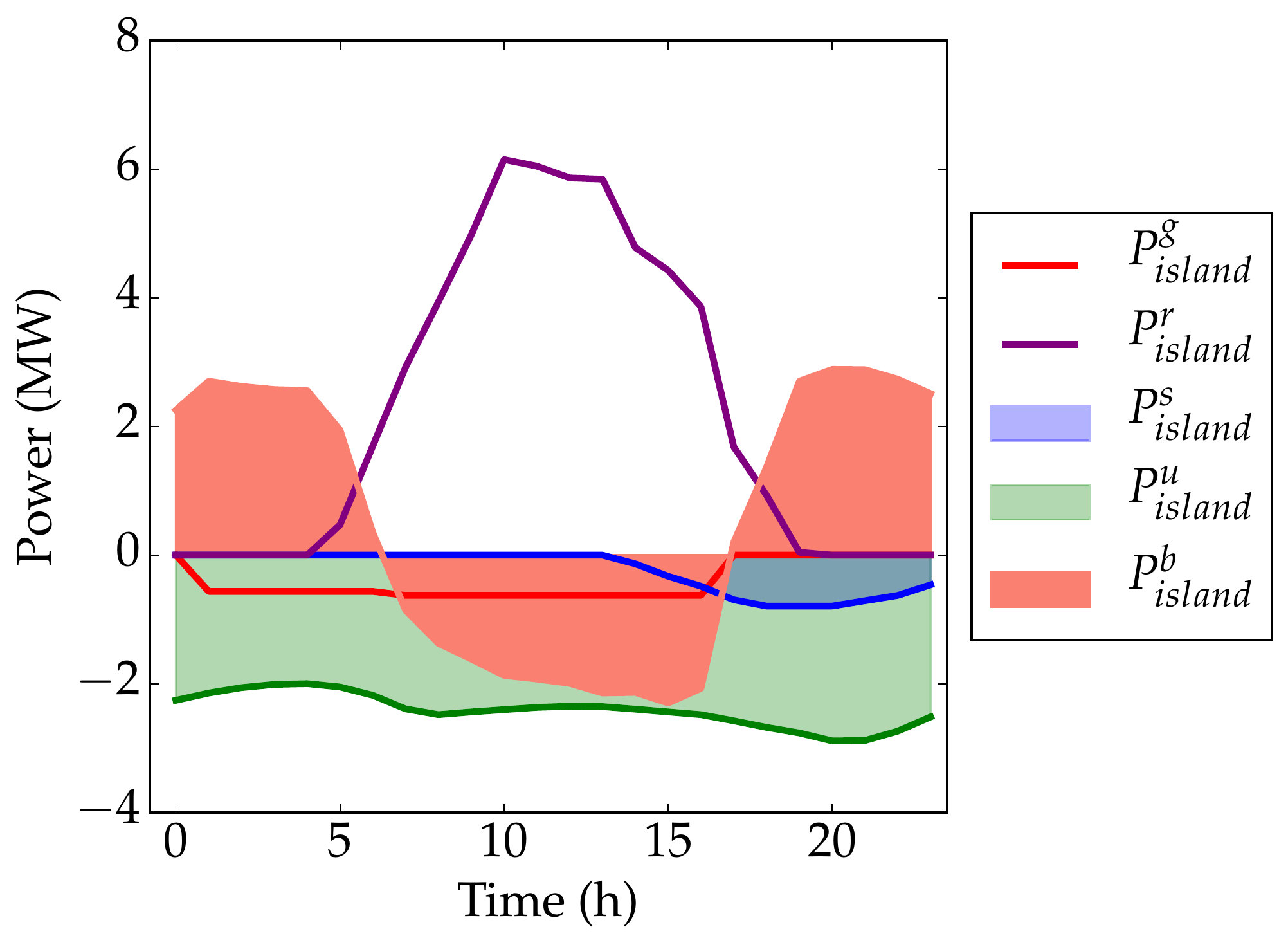}
	\caption{Results for an islanded, autonomous microgrid operation.}
	\label{fig:island}
	    \vspace{-0.5cm}
\end{figure}
Fig.~\ref{fig:sumofnetwork_CESID} depicts the aggregate load profile and generation after Algorithm~\ref{cecalgo} terminates. More renewable energy is injected into the grid and traded via the designed incentives for $\mc{CT}_2$ crowdsourcees. The net contribution of $\mc{CT}_2$ crowdsourcees is shaded in red. It is noteworthy to mention that the utility cannot schedule the shapeable loads of $\mc{CT}_2$ crowdsourcees. The blue area in Fig.~\ref{fig:sumofnetwork_CESID} displays the unexpected load demand of $\mc{CT}_2$ crowdsourcees. The generator at the substation covers this demand shortage; see Fig.~\ref{fig:sumofnetwork_CESID} where $P^g_{\mathrm{CES-ID}}$ is greater than $P^g_\mathrm{CES-OPF}$ from 3 pm to 11 pm.
  
\subsubsection{Islanded Microgrid Test}~\label{sec:islandedtest}
After implementing P2P energy trading, we simulate a scenario of a small islanded, autonomous microgrid. In this microgrid, we assume the following. First, all users have (a) enough solar power to produce enough energy to supply the grid, and (b) the microgrid has a battery with sufficient capacity to store excess solar energy. Second, each user agrees to participate in the program and their DERs would be fully controlled by the microgrid management algorithm akin to Algorithm~\ref{cecalgo}. The simulation setup remains the same as in Section~\ref{sec:Setup} except the solar panels produce more energy and the capacities of batteries are enlarged. Fig.~\ref{fig:island} shows the outcome of the autonomous microgrid operation. Between 6 am and 7 pm, the solar panel on each crowdsourcees' roof not only produces enough energy to meet the real-time load demand but also stores excess energy into batteries for night use. At night, batteries start to discharge energy to cover the demand shortage facilitating energy trading transactions with crowdsourcees in need for energy using blockchain and smart contracts. 
 \subsubsection{Blockchain and ETT GUI}
Fig.~\ref{fig:GUI.png} shows a web-based user prototype that we implemented using Hyperledger Fabric as described in Section~\ref{sec:BlockChain}. The web application shows the system operation which includes creating crowdsourcees, selling energy to the utility or neighborhood, and listing all energy trading transactions with information about the prices and the users. This web-based prototype interacts with the optimization solvers and algorithms that generate forecasts, as well as the crowdsourcees. 
\begin{figure}[t]
    \centering \includegraphics[scale=0.15]{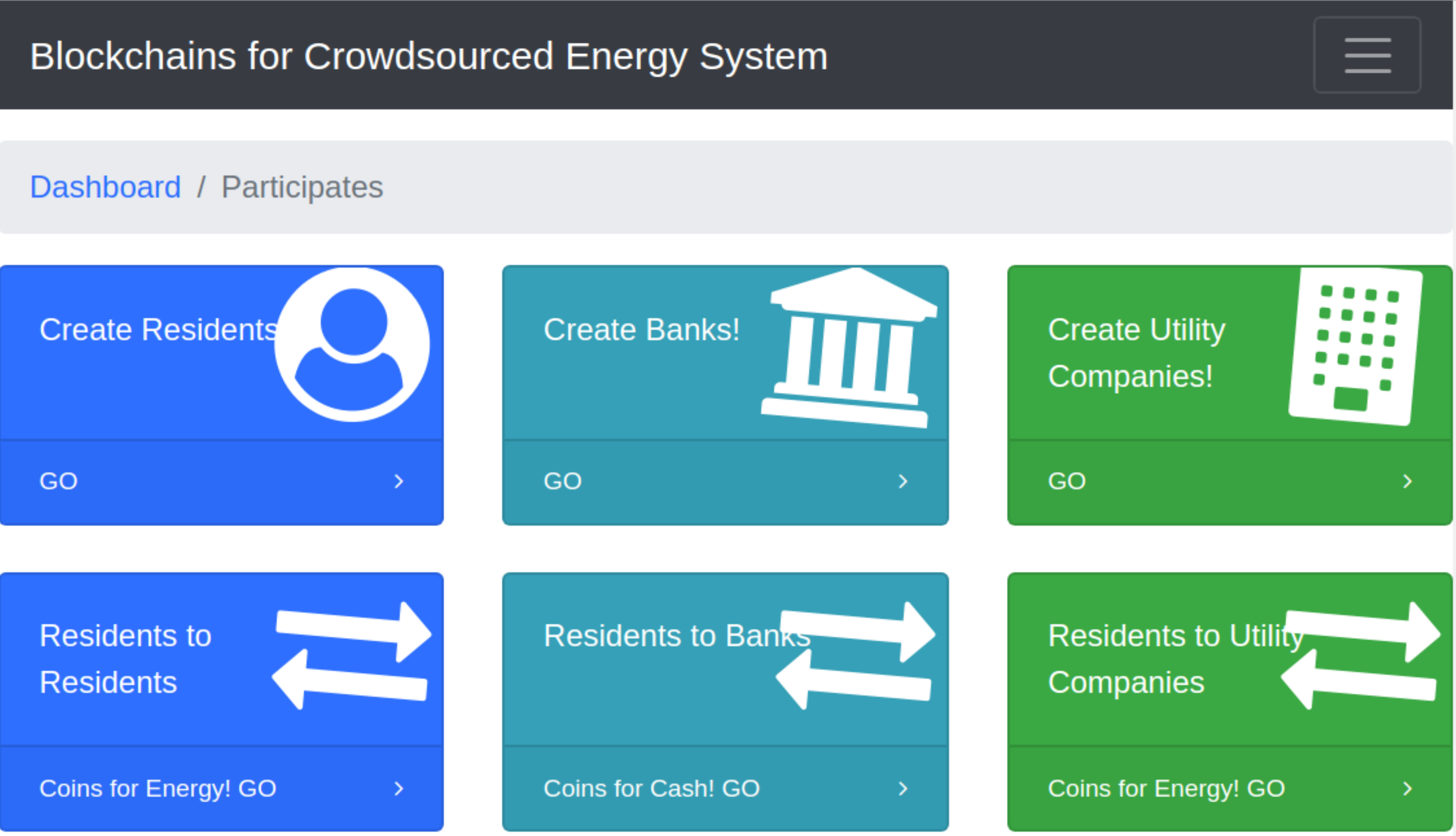}
    \centering \includegraphics[scale=0.15]{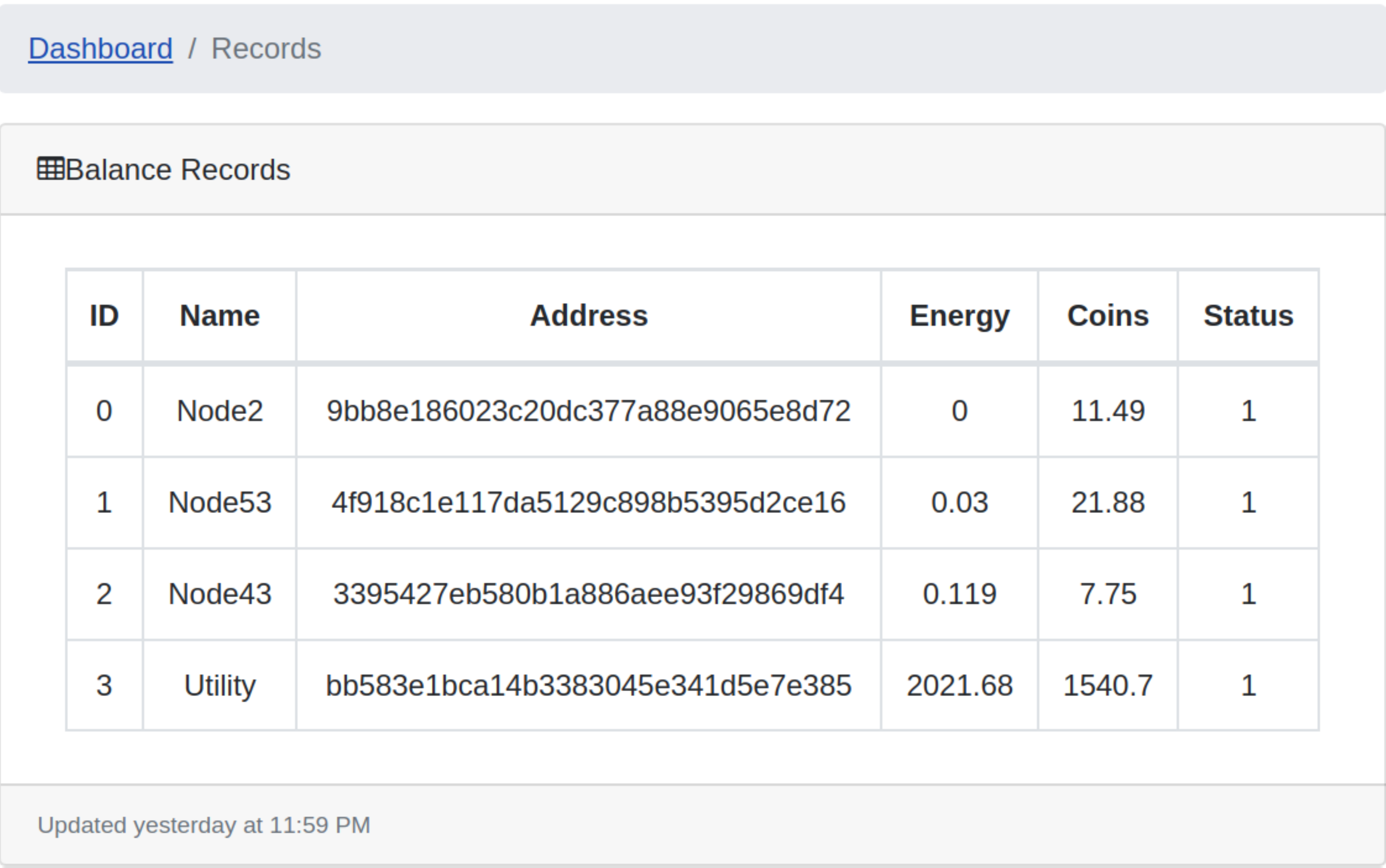}
    \centering \includegraphics[scale=0.15]{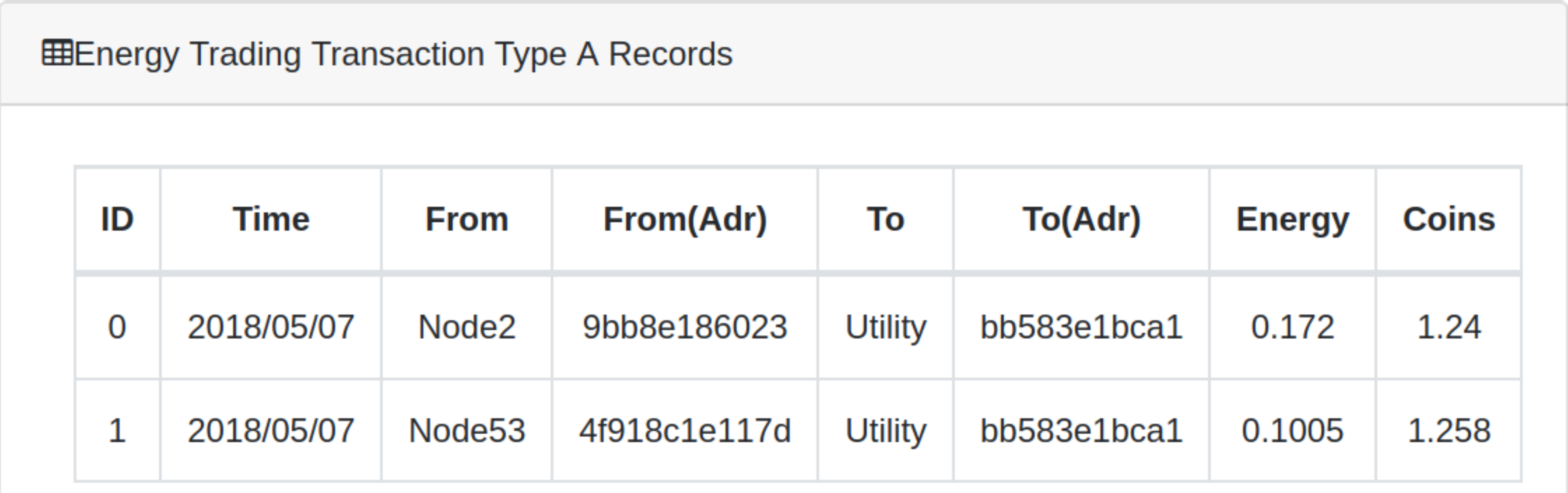}
    \centering \includegraphics[scale=0.15]{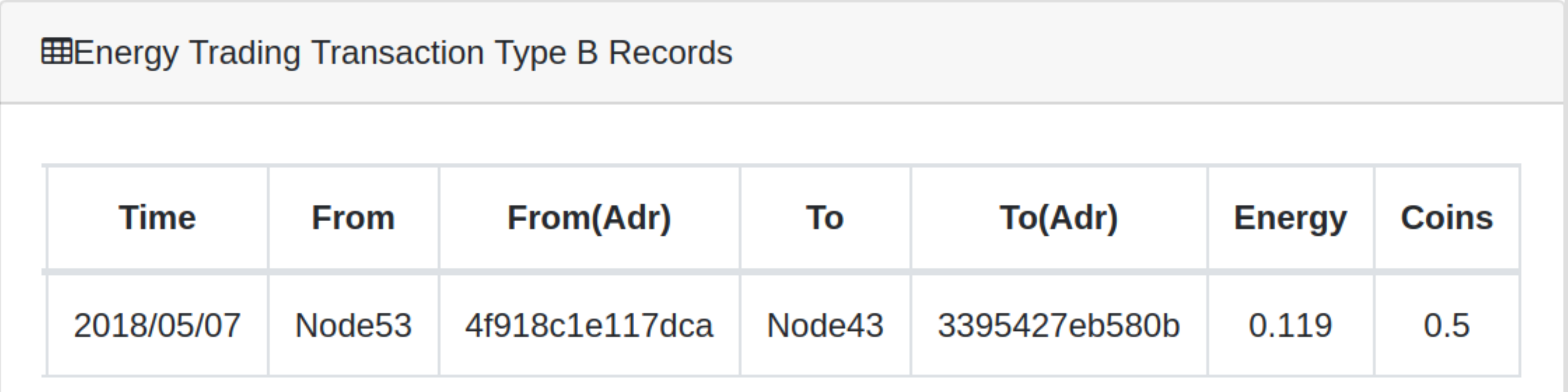}
    \caption{Web-based user interface for CESs with Hyperledger Fabric.}
    \label{fig:GUI.png}
    \vspace{-0.5cm}
\end{figure}
\section{Paper Summary, Limitations and Future Work}
The paper introduces the notion of blockchain-assisted crowdsourced energy systems with a specific implementation and prototype of blockchain that scales to include millions of crowdsourcees and P2P energy trading transactions. A thorough review of the blockchain technology for energy systems is given. Various types of crowdsourcees and energy trading transactions are introduced to mimic current and projected energy market setups. Then, an operational OPF-based model of CESs with batteries, shapeable loads, and other DERs is introduced for distribution networks---considering energy trading transactions and crowdsourcees preferences---yielding a day-ahead market equilibrium. Monetary incentives are designed to attract crowdsourcees in hour-ahead and real-time markets to the computed equilibrium while satisfying a demand shortage or surplus. Furthermore, an implementation of blockchain through the IBM Hyperledger Fabric is discussed with its coupling with the optimization models. This implementation allows the system operator to manage the network users to seamlessly trade energy.  Finally, case studies are given to illustrate the practicality of the presented methods for classical distribution networks, as well as self-sufficient and islanded microgrids.

There is still a uncontrollable risk in blockchain based energy trading system, i.e., the attack from malicious market operator, stakeholders or outsider. \textit{(1)} A malicious market operator will attempt to modify the operation of the market algorithms in order to produce results that provide an output (market price, load demand) providing them with a financial advantage over the authentic price or demand outputs. \textit{(2)} A malicious stakeholder might try to  produce a false clearing price offering them with reduced energy costs. However, because Hyperledger messages are digitally signed, the consensus results will not be manipulated as long as there are $2f+1$ total operators, where $f$ is the number of malicious operators. \textit{(3)} A malicious outsider will try to remotely tamper with all messages communicated between the crowdsourcees and market operators. Their goal is to manipulate the resulting market operations in order to either manipulate \textit{(i)} the load demand bids submitted by the crowdsourcees or \textit{(ii)} the market clearing prices.
 
In future work, we plan to extend the presented research to address distributed consensus mechanisms for blockchain in crowdsourced energy system, and threats from malicious crowdsourcees, market operators, and outsiders. 

\tiny
\bibliographystyle{IEEEtran}
\bibliography{IEEEabrv,bibfile}

\vskip -3\baselineskip plus -1fil
\begin{IEEEbiography}
	[{\includegraphics[width=1in,height=1.25in,clip,keepaspectratio]{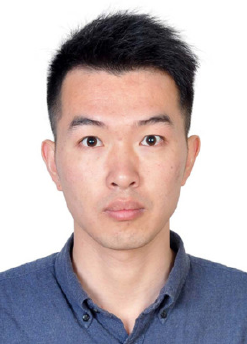}}]
	{Shen Wang}  received the master's degree in Control Science and Engineering from the University of Science and Technology of China, Hefei, China, in 2016. He is currently pursuing a Ph.D. degree in Electrical Engineering at the University of Texas at San Antonio, Texas. His current research interests include optimal control in cyber-physical systems with special focus on energy and water systems.
\end{IEEEbiography} 

\vspace{-0.40cm}
\vskip -3\baselineskip plus -1fil

\begin{IEEEbiography}
	[{\includegraphics[width=1in,height=1.25in,clip,keepaspectratio]{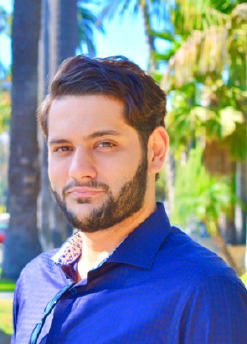}}]
	{Ahmad F. Taha} is  an assistant professor with the Department of Electrical and Computer Engineering at the University of Texas, San Antonio. He received the B.E. and Ph.D. degrees in Electrical and Computer Engineering from the American University of Beirut, Lebanon in 2011 and Purdue University, West Lafayette, Indiana in 2015.  Dr. Taha is interested in understanding how complex cyber-physical systems (CPS) operate, behave, and \textit{misbehave}. His research focus includes optimization, control, and security of CPSs with applications to power, water, and transportation networks.
\end{IEEEbiography}
\vspace{-0.40cm}
\vskip -3\baselineskip plus -1fil

\begin{IEEEbiography}
	[{\includegraphics[width=1in,height=1.25in,clip,keepaspectratio]{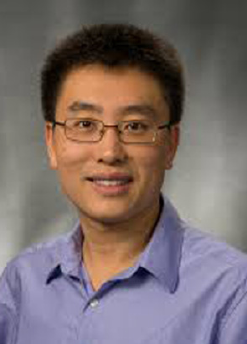}}]{Jianhui Wang (M'07-SM'12)}
received the Ph.D. degree in electrical engineering from Illinois Institute of Technology, Chicago, Illinois, USA, in 2007.
Presently, he is an Associate Professor with the Department of Electrical Engineering at Southern Methodist University, Dallas, Texas, USA. Prior to joining SMU, Dr. Wang had an eleven-year stint at Argonne National Laboratory with the last appointment as Section Lead – Advanced Grid Modeling. Dr. Wang is the secretary of the IEEE Power \& Energy Society (PES) Power System Operations, Planning \& Economics Committee. He has held visiting positions in Europe, Australia and Hong Kong including a VELUX Visiting Professorship at the Technical University of Denmark (DTU). Dr. Wang is the Editor-in-Chief of the IEEE Transactions on Smart Grid and an IEEE PES Distinguished Lecturer. He is also a Clarivate Analytics highly cited researcher for 2018.
\end{IEEEbiography} 

\vspace{-0.40cm}
\vskip -3\baselineskip plus -1fil

\begin{IEEEbiography}
	[{\includegraphics[width=1in,height=1.25in,clip,keepaspectratio]{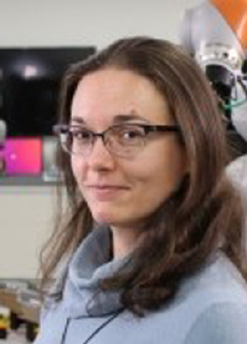}.pdf}]
	{Karla Kvaternik}  is Research Scientist in the Predictive Analytics research group at Siemens Corporate Technology, in Princeton, NJ. Dr. Kvaternik holds a Ph.D. in systems and control theory from the University of Toronto, focusing on the synthesis of decentralized optimization algorithms, networked multiagent coordination control methods and stability analysis techniques thereof. Prior to her doctoral studies, she won the Best Student Paper Award at the IEEE Multiconference on Systems and Control, for her M.Sc. work on multivariable output feedback for nonlinear systems. As a Research Associate at Princeton University, Karla studied multiagent reinforcement learning algorithms applied to multiarmed bandit settings. At Siemens, Dr. Kvaternik enjoys addressing a variety of data-driven decision support problems and prototyping transactive energy applications.
\end{IEEEbiography} 
\vspace{-0.40cm}
\vskip -3\baselineskip plus -1fil

\begin{IEEEbiography}
	[{\includegraphics[width=1in,height=1.25in,clip,keepaspectratio]{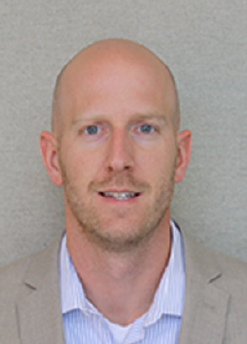}}]
	{Adam Hahn}  is currently an assistant professor in the Department of Electrical Engineering and Computer Science at Washington State University. His research interests include cybersecurity of the smart grid and cyber-physical systems (CPS), including intrusion detection, risk modeling, vulnerability assessment, and secure system architectures. He received M.S. and Ph.D. degrees from the Department of Electrical and Computer Engineering at Iowa State University in 2006 and 2013. Previously, he worked as a Senior Information Security Engineer at the MITRE Corporation, supporting numerous cybersecurity assessments within the federal government and leading research projects in CPS security
\end{IEEEbiography}

\end{document}

%% file: preambleT.tex
\usepackage{epsfig,color,amsmath,cite}
\usepackage{wrapfig}
\usepackage{bm}
\usepackage{epstopdf}
\usepackage{amssymb}
\usepackage{url}
\usepackage{enumitem}
\usepackage{multirow}
\usepackage{booktabs}
\usepackage{makecell}
\usepackage{algorithm,algorithmic}
\setlist[itemize]{leftmargin=*}

\makeatother

\newcommand{\bmat}[1]{\begin{bmatrix} #1 \end{bmatrix}}

\newcommand{\m}{\boldsymbol}
\allowdisplaybreaks[4]

\newcommand{\mc}[1]{\mathcal{#1}}

\usepackage[framemethod=TikZ]{mdframed}

\mdfdefinestyle{MyFrame}{%
	linecolor=black,
	outerlinewidth=1pt,
	roundcorner=4pt,
	innerrightmargin=5pt,
	innerleftmargin=5pt,}

\usepackage{pifont}
\newcommand{\pie}[1]{%
	\begin{tikzpicture}
	\draw (0,0) circle (0.7ex);\fill (0.7ex,0) arc (0:#1:0.7ex) -- (0,0) -- cycle;
	\end{tikzpicture}%
}

\usepackage[flushleft]{threeparttable}